\documentclass[11pt]{article}
\usepackage[margin=1in]{geometry}
\usepackage{amsmath,amssymb}
\usepackage{graphicx}
\graphicspath{{./}}
\usepackage{natbib}
\usepackage{hyperref}
\usepackage{booktabs}
\usepackage{setspace}
\usepackage{lineno}
\usepackage{xcolor}

\title{Conditional Tropical Cyclogenesis Rates via Rare-Event Sampling in a Neural Weather Emulator}

\author{%
  John S. Schreck$^{1,*}$,
  William Chapman$^{2}$, 
  Charlie Becker$^{1}$,
  David John Gagne II$^{1}$ \\
  \small $^1$NSF National Center for Atmospheric Research, Boulder, CO \\
  \small $^2$University of Colorado Boulder, Boulder, CO \\
  \small $^*$Corresponding author: schreck@ucar.edu
}

\date{\today}

\begin{document}
\maketitle

\begin{abstract}
We couple Forward Flux Sampling (FFS), a non-equilibrium rare-event technique from statistical mechanics, to a neural weather emulator (SDL-WXFormer, 1$^\circ$ grid spacing) to estimate conditional tropical cyclogenesis rates, or how often a tropical cyclone achieves a hurricane-level central pressure, without modifying model dynamics. Tropical cyclogenesis rates vary by orders of magnitude across regimes, yet direct ensemble sampling cannot resolve this variability at operationally feasible ensemble sizes. FFS decomposes the rare disturbance to mature cyclone intensification path into a flux through an initial interface pressure and a product of conditional crossing probabilities across four intermediate  interface pressures. We use the 1$^\circ$ emulator because FFS requires $\mathcal{O}(10^4)$ model trajectories per initial condition, and because the model's calibrated stochastic layers provide the necessary exploratory spread. Applied to 98 Atlantic basin initial conditions spanning 21~August -- 8~October~2022, FFS resolves genesis rates spanning nearly three orders of magnitude, capturing a  seasonal cycle qualitatively consistent with observations. A self-consistency check comparing FFS rates to independent direct-sampling rates yields a mean ratio of $1.03 \pm 0.15$ across all initial conditions. Computational enhancement factors range from $3\times$ (most active environment) to $140\times$ (most suppressed), with a geometric mean of $14\times$. Three case studies illustrate the physical diagnostics the method provides: the rate-limiting step is initial tropical organization for the Earl environment, uniformly high crossing probabilities for the Fiona precursor environment, and a compound barrier at the final intensification stages for the Ian environment. More efficient emulators would enable application of FFS to finer resolutions.
\end{abstract}

\section{Introduction}

Tropical cyclogenesis, or the process of a tropical disturbance intensifying into a mature tropical cyclone, represents one of the most consequential rare events in atmospheric science. Approximately 10--15\% of Atlantic tropical disturbances develop into named systems \citep{Frank1970, gray1968global, Landsea1998}, but discriminating developing from non-developing cases challenges deterministic and ensemble approaches alike
\citep{Titley2000, Halperin2013}. The challenge is not merely predictive: it is statistical. For a genesis event with probability $p \sim 10^{-3}$ per day, direct Monte Carlo estimation of the genesis rate to 20\% relative uncertainty requires $\mathcal{O}(10^4)$ ensemble trajectories per initial condition, far beyond both operational and research budgets when using physics-based models. The question of \textit{how frequently} genesis occurs under a given atmospheric state, not just whether individual ensemble members develop storms, remains largely inaccessible to
conventional approaches.

Rare event sampling techniques address this by concentrating computation on the transition itself. In atmospheric science, most work has used genealogical large-deviation algorithms \citep{ragone2018computation, galfi2019large}, Adaptive Multilevel Splitting \citep[AMS;][]{cerou2007adaptive}, and related trajectory-resampling methods \citep{webber2019practical, finkel2023revealing, finkel2024bringing}, including applications to tropical cyclone intensification \citep{plotkin2019maximizing}. These methods yield probability and return-time estimates in addition to rare event realizations, and recent work has begun coupling them to AI emulators \citep{lancelin2025ai, manshausen2026extreme}. Forward Flux Sampling \citep[FFS;][]{allen2006forward, allen2009forward, valeriani2007computing, hussain2020studying} is complementary. It computes a transition rate constant by factorizing the transition across a fixed ladder of interfaces, and the per-interface conditional probabilities show which stage of the transition is hardest to cross. That staged decomposition is the property we exploit here, since it identifies which precursor processes are most critical for genesis to occur in a given environment.

FFS requires no equilibrium assumptions: no free energy landscapes, detailed balance, or Boltzmann distributions. The method assumes only (1) well-defined metastable regions, (2) measurable flux through interfaces, and (3) Markovian dynamics \citep{allen2009forward}.
It has been applied across a broad range of non-equilibrium systems from biochemical networks \citep{allen2005sampling} to nucleic acid folding kinetics \citep{schreck2015dna}. The atmosphere is a canonical non-equilibrium steady-state (NESS) system driven persistently by solar forcing and dissipated by radiation, friction, and turbulence; and satisfies all three above requirements. Atmospheric rare events are dynamical transitions between regions of a strange attractor \citep{lorenz1969atmospheric,touchette2009large}, not thermodynamic transitions over energy barriers, making FFS well suited to this setting.

FFS genesis rates vary by three orders of magnitude across the 98 initial conditions studied here, reflecting strongly non-stationary
background conditions across the hurricane season. The quantity $k$ computed for each initial condition is a \textit{conditional genesis rate given the atmospheric state at that IC}, not a climatological steady-state rate. Each FFS simulation treats a
10--15~day forecast window anchored to a specific ERA5 state as locally stationary, so $k$ characterizes genesis probability density in that particular atmospheric environment. This conditional framing is analogous to a conditional return period
given present climate state \citep{finkel2023revealing, finkel2024bringing}. It motivates our seasonal sampling across 98~ICs and enables direct comparison against observed
Atlantic activity.

Machine learning weather models \citep{lam2023graphcast, bi2023pangu, pathak2022fourcastnet}
make this calculation practical because they are much faster than physics-based models. That speed matters here because FFS requires $\mathcal{O}(10^4)$ trajectories per initial condition. SDL-WXFormer \citep{schreck2025sdl} is useful for a second reason: its stochastic decomposition layers produce the calibrated ensemble spread required by the shooting, or conditional stochastic generation, phase. 

This paper asks a simple question: can FFS recover useful conditional genesis rates from a fast weather emulator, and if so, what parts of genesis control those rates? We deliberately use SDL-WXFormer at 1$^\circ$ grid spacing rather than a convection-permitting physics-based model for two practical reasons. First, FFS is computationally demanding: each initial condition requires $\mathcal{O}(10^4)$ 15-day trajectories, and at $\approx$0.24 GPU-minutes per single-member 15-day forecast on an NVIDIA A100, the full 98-IC dataset required approximately 2,450 GPU-hours. A comparable experiment with WRF or IFS would be prohibitively expensive on available allocations. Second, the SDL-WXFormer's calibrated stochastic decomposition layers provide the ensemble variability the FFS shooting phase requires. A deterministic neural emulator would collapse all branching probabilities to zero or one. The 1$^\circ$ grid spacing imposes a well-understood limitation: the model cannot faithfully reproduce Category~3+ intensity or convective-scale vortex dynamics, so state~$B$ is defined at 975~hPa (tropical storm to minimal hurricane) rather than the intensities typical of major hurricane development. That limitation matters for the threshold definition, but not for the main point here, which is whether the FFS calculation is internally consistent and physically informative.

Whether the FFS rate calculation is \textit{physically} Markovian depends on whether
the model state includes the variables that matter for genesis.
SDL-WXFormer does not explicitly predict slow boundary-condition fields such as
sea-surface temperature (SST) or soil moisture. These are held fixed at their initialization values rather than evolving during the simulation, though the model's weights encode multi-step climatological relationships between these slow variables and atmospheric dynamics learned during training. This can introduce background-state dependence that a fully Markovian FFS calculation
would not have. The self-consistency check described in Section~\ref{sec:results_val} can detect Markov violations that produce systematic bias in the factorized rate relative to the direct count, but cannot detect violations that bias both estimators in the same direction. We defer a more rigorous test of physical Markovianity to future work.

This paper reports: (1) a three-order-of-magnitude seasonal cycle in conditional genesis
rates qualitatively consistent with the known structure of the 2022 Atlantic season; (2) a self-consistency validation showing FFS and direct-sampling rates agree within a mean ratio of $1.03\pm0.15$ across 98 initial conditions; (3) case studies illustrating physically distinct rate-limiting steps; and (4) a Cyclone Phase Space \citep[CPS;][]{hart2003cyclone} contamination screen separating tropical from extratropical development pathways.

\section{Methods}

\subsection{Forward Flux Sampling}
\begin{figure}[htbp]
\centering
\includegraphics[width=\textwidth]{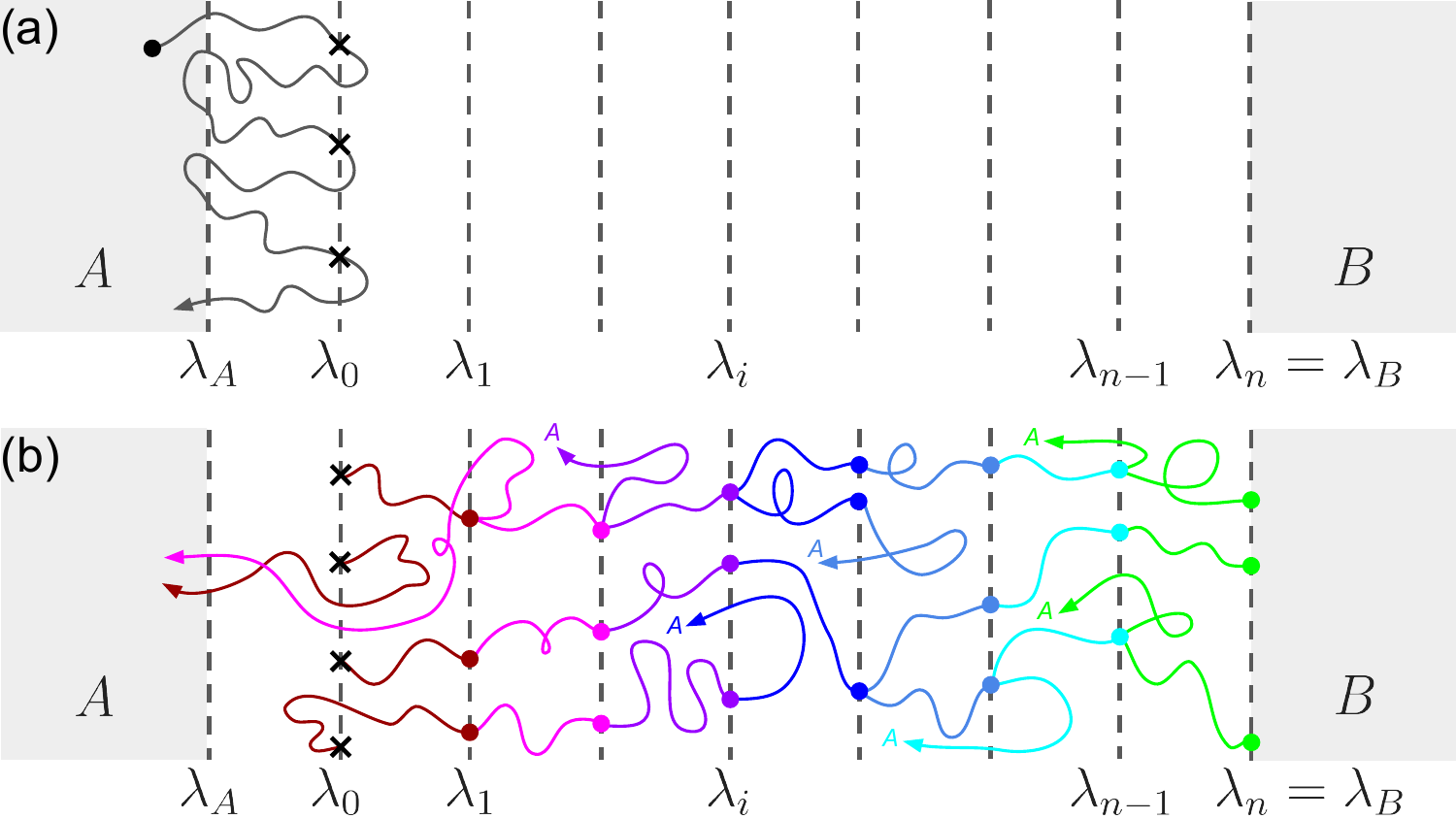}
\caption{Schematic of the Forward Flux Sampling algorithm (after \citealt{allen2006forward}).
\textit{Panel (a), flux phase}: Gray trajectories evolve forward from state $A$.
Each decorrelated crossing of $\lambda_0$ is recorded (marked $\times$).
\textit{Panel (b), shooting phase}: New trial trajectories are launched from each saved
$\lambda_i$ configuration (filled circles). Shots that return to state $A$ (arrows labeled $A$)
are failures; shots that reach $\lambda_{i+1}$ (filled circles on the next interface) are successes,
yielding the conditional probability $P(\lambda_{i+1}|\lambda_i)$.
}
\label{fig:schematic}
\end{figure}
FFS estimates the transition rate from metastable state $A$ (unorganized disturbances, $Q > \lambda_A$) to state $B$ (organized tropical cyclone, $Q \leq \lambda_B$) by factorizing:
\begin{equation}
k_{A\to B}^{\text{FFS}} = \Phi_{A,0} \times \prod_{i=1}^{n} P(\lambda_i | \lambda_{i-1}, \mathbf{x_{\Omega}})
\label{eq:ffs_rate}
\end{equation}
where $\Phi_{A,0} = N_{\text{cross}}/T_{\text{total}}$ is the flux of trajectories crossing the first interface $\lambda_0$ per unit time, and $P(\lambda_i|\lambda_{i-1})$ is the conditional probability that a trajectory at $\lambda_{i-1}$ reaches $\lambda_i$ before returning to $A$ (Fig.~\ref{fig:schematic}). The conditioning on $\mathbf{x_{\Omega}}$, the atmospheric state over the basin domain $\Omega$ at the initial condition, makes explicit that each rate is specific to its initialization. The order parameter $Q(\mathbf{x})$ is the minimum mean sea-level pressure (MSLP) within the Atlantic basin domain $\Omega = [10^\circ\text{N},\,40^\circ\text{N}] \times [100^\circ\text{W},\,20^\circ\text{W}]$ derived from atmospheric state $\mathbf{x}$.

We use MSLP because it is one of the standard tropical cyclone intensity measures and because it can be tracked cleanly on this grid \citep{knaff2007improved}. MSLP provides a downstream summary of the thermodynamic and kinematic precursors to genesis. FFS requires that the chosen interfaces separate the
transition in a physically sensible way. Based on our analysis, they do: the seasonal cycle in
Figure~\ref{fig:rates} is coherent, and the state-$B$ arrivals in
Figure~\ref{fig:stateb_geo} cluster in the main development region.
Future implementations could extend the order parameter definition to incorporate low-level vorticity,
mid-tropospheric moisture, or a composite genesis potential index
\citep[e.g.,][]{bister1998dissipative,camargo2007use}.
Such extensions may sharpen the bottleneck diagnostics, but they are not needed to test the rate calculation used here.

We define State $A$: $Q > \lambda_A = 1008$~hPa, a pressure generally associated with unorganized synoptic disturbances.
For State $B$ we expect both $Q \leq \lambda_B = 975$~hPa and that the warm-core structure is satisfied based on the CPS criterion and that the vortex center is within the basin domain $\Omega$.
We note that 975~hPa corresponds to tropical storm to minimal hurricane intensity and is \textit{intentionally stronger} than the conventional NHC tropical storm genesis designation
($\sim$34~kt, roughly 997--1003~hPa in typical Atlantic pressure--wind relationships).
At 1$^\circ$ grid spacing, the model's smoothed MSLP field does not develop a reliable closed
isobar signature until the vortex is more strongly organized than the standard genesis threshold.
In practice, 975~hPa is the lowest pressure at which state-$B$ arrivals cluster in physically expected
geographic locations and pass the CPS warm-core screen with consistent reliability.
The compound criterion means that reaching the 975~hPa MSLP threshold is necessary but not sufficient
for state~$B$. Trajectories that deepen to 975~hPa while recurving poleward or exiting the domain
are not counted as state-$B$ arrivals, and $P_4 = P(B|\lambda_3) < 1$ reflects this screening
in addition to the difficulty of the final pressure crossing.
The 975~hPa threshold is calibrated to the SDL-WXFormer's 1$^\circ$ horizontal
grid spacing through trial experiments (see Appendix~\ref{sec:si_threshold}).
The flux interface and four shooting interfaces are:
\begin{equation*}
\lambda_0 = 1000.0,\quad
\lambda_1 = 987.2,\quad
\lambda_2 = 984.7,\quad
\lambda_3 = 981.2,\quad
\lambda_4 = \lambda_B = 975.0 \quad [\text{hPa}]
\end{equation*}
satisfying $\lambda_A > \lambda_0 > \lambda_1 > \cdots > \lambda_4 = \lambda_B$.
The interfaces and thresholds stated here are those used to produce all results in this paper
and match the configuration file in the public code repository
(\texttt{config/ffs.yml}; \url{https://github.com/NCAR/miles-tails}).
Note that \texttt{forecast\_start\_times} in that file lists only the three case-study ICs.
The full 98-IC seasonal run was launched via the PBS job scripts in
\texttt{applications/run\_paper\_figures\_mar18.sh}.

We label the four conditional crossing probabilities $P_1$--$P_4$ in order of the
FFS product: $P_1 = P(\lambda_1|\lambda_0)$, $P_2 = P(\lambda_2|\lambda_1)$,
$P_3 = P(\lambda_3|\lambda_2)$, and $P_4 = P(B|\lambda_3)$.
Configurations saved at $\lambda_0$ during the flux phase seed the first shooting
step, so $P_1$ measures the first interface crossing
(1000.0 $\to$ 987.2~hPa) and $P_4$ measures the final crossing from
$\lambda_3$ (981.2~hPa) into state~$B$.

\begin{figure}[htbp]
\centering
\includegraphics[width=\textwidth]{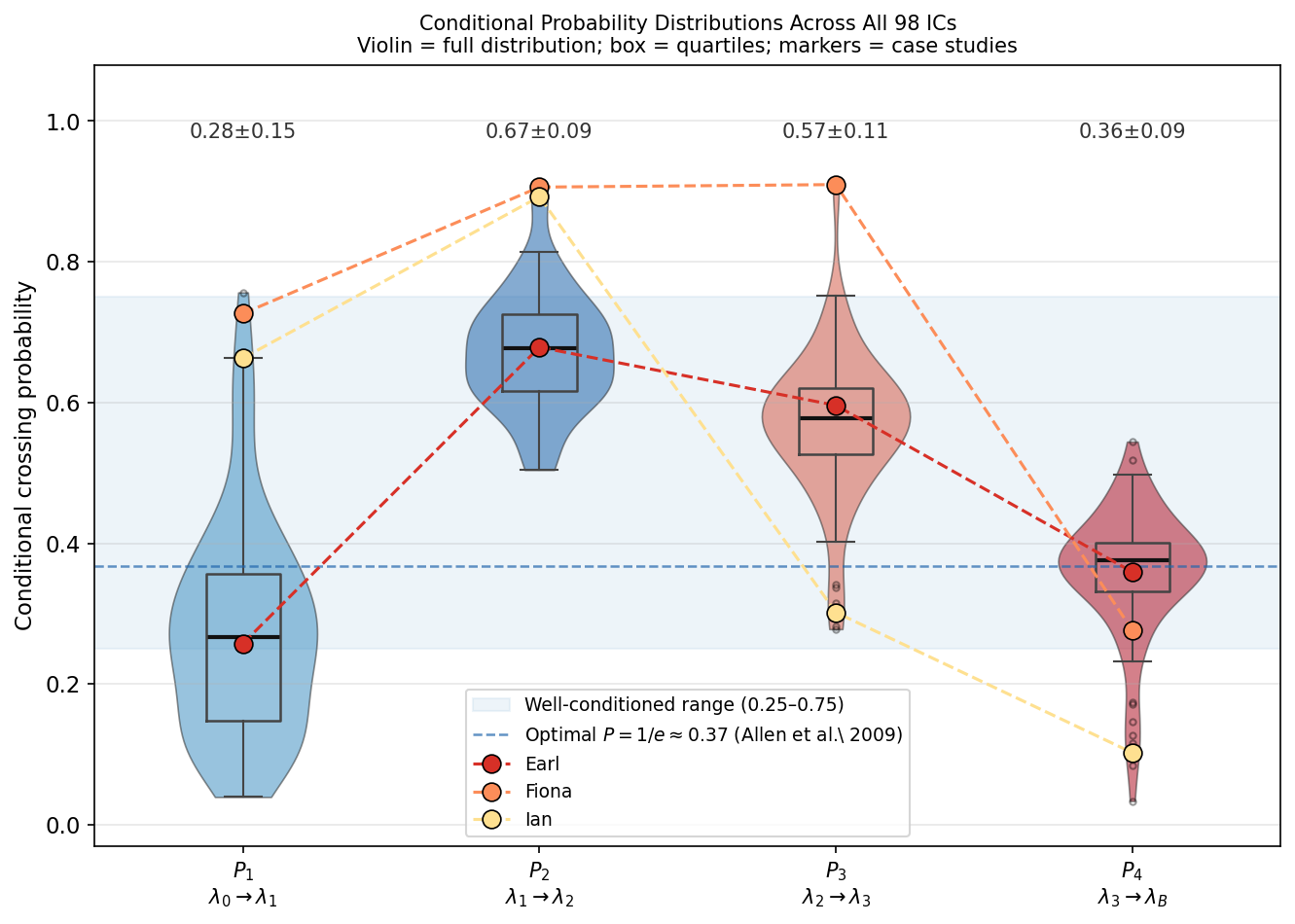}
\caption{Distribution of conditional crossing probabilities $P_i$ across all 98 initial
conditions. Each violin shows the full distribution; boxes show quartiles; coloured
lines connect the three case-study ICs. The horizontal dashed line marks the optimal
$P = 1/e \approx 0.37$ \citep{allen2009forward}; the blue band spans the well-conditioned range 0.25--0.75.
Most ICs have $P_i$ within the well-conditioned range, confirming that interface placement
produces efficient variance reduction. The spread of $P_1$ (initial organization) is larger
than that of $P_2$--$P_3$, reflecting the dominant role of large-scale atmospheric variability
in setting the first intensification barrier. Mean $\pm$ SD for each interface are annotated.}
\label{fig:cond_prob_dist}
\end{figure}
Shooting interfaces are chosen using the optimal-$1/e$ rule of
\citet{allen2009forward}: the next interface $\lambda_{i+1}$ is placed at the
$p^*$-quantile of the MSLP distribution reached by trajectories launched from
$\lambda_i$, where $p^* = 1/e \approx 0.368$.  This placement minimizes variance in
the estimated rate and targets conditional crossing probabilities near $1/e$,
maximizing statistical efficiency \citep{allen2009forward}.
In practice, interfaces were optimized from long trial runs and then rounded to physically interpretable pressure values; the resulting $P_i$ values (Table~\ref{tab:results})
cluster between 0.10 and 0.91 across all ICs, with a majority in the well-conditioned
range 0.25--0.75 as shown in Figure~\ref{fig:cond_prob_dist}.

For each initial condition we generate trajectories from the ERA5 state perturbed by
SDL-WXFormer stochastic layers with 6-hour prediction time steps. New storm tracks are seeded for
up to 15 days. Any active storms at that point continue to be followed until they
return to state $A$ or reach state $B$. Decorrelated
$\lambda_0$ crossings (requiring return to state $A$ between consecutive crossings) are
accumulated to $N_{\text{cross}} = 2000$ per initial condition, yielding Poisson
uncertainty $\sigma_\Phi/\Phi \approx 2.2\%$, well below across-IC variability.
State-$B$ arrivals during flux generation provide an independent direct-sampling rate
$k^{\text{direct}} = N_{\text{direct}}/T_{\text{total}}$ that serves as a self-consistency check.

Configurations saved at $\lambda_0$ seed shooting trajectories toward each successive
interface. For each interface $\lambda_{i-1}$, trajectories are launched until 2000
successful crossings of $\lambda_i$ are accumulated (Appendix~\ref{sec:convergence} justifies this target). Success probability
$P(\lambda_i|\lambda_{i-1}) = N_{\text{success}}/N_{\text{attempts}}$ with binomial
uncertainty. Statistical uncertainty on the full rate (Eq.~\ref{eq:ffs_rate}) propagates
from each factor:
\begin{equation}
\frac{\sigma_k}{k} = \sqrt{\left(\frac{\sigma_\Phi}{\Phi}\right)^2 + \sum_{i=1}^{n}\left(\frac{\sigma_{P_i}}{P_i}\right)^2}
\end{equation}

A trial trajectory that successfully crosses all four shooting interfaces from $\lambda_0$
through to state~$B$ is called a \textit{reactive pathway}.
From a single $\lambda_0$ configuration, each successful crossing at interface $\lambda_i$
seeds new launches toward $\lambda_{i+1}$, producing a branching genealogy called a
\textit{shooting tree}.
Multiple reactive pathways that descend from the same $\lambda_0$ seed carry a
\textit{cluster weight} (branching factor) proportional to the statistical weight
of that seed in the FFS ensemble. Pathways from seeds that barely cleared each
interface are weighted lower accordingly.
The cumulative product $p_B(\lambda_i) = \prod_{j \geq i} P_j$ is the
\textit{committor}: the probability that a trajectory arriving at $\lambda_i$ eventually
reaches state~$B$ \citep{bolhuis2002transition, Khoo2019}.

\subsection{SDL-WXFormer and Experimental Setup}

SDL-WXFormer \citep{schreck2025sdl} is a hierarchical encoder-decoder vision transformer
with U-Net skip connections, trained on ERA5 reanalysis \citep{hersbach2020era5} (1979--2021,
test year 2022) at 1$^\circ$ grid spacing and 6-hour timesteps.
The model predicts 64 upper-air channels (U, V, T, Q at 16 ERA5 model levels) plus 7 surface
variables, and accepts 3 input-only forcing channels (solar, orography, land mask).
Stochastic Decomposition Layers (SDL) inject three independent latent noise vectors
into the decoder at three spatial scales (after each upsampling stage, before
skip-connection concatenation), producing calibrated ensemble spread through 15-day
lead times.
A single 15-day forecast (60 steps) requires $\approx$0.24 GPU-minutes on a single
NVIDIA A100. Scaling to 10 ensemble members, whose trajectories are batched through
the model in parallel, requires $\approx$1.96 GPU-minutes.
Enabling rare-event sampling via FFS was an explicit design objective of SDL-WXFormer.
SDL-WXFormer operates as a strict state-in/state-out emulator: each 6-hour step
takes only the current atmospheric state vector as input and applies the learned weights
plus an independent stochastic perturbation drawn from the SDL layers, with no recurrent
memory carried between steps.
This design makes SDL-WXFormer Markovian at inference by construction, satisfying the
core FFS requirement.

We select 98 initial conditions from ERA5 at 00 and 12~UTC spanning August~21 --
October~8, 2022, verifying $Q(\mathbf{x}_0) > 1008$~hPa for each. Three initial
conditions receive detailed case study treatment:
\begin{itemize}\setlength\itemsep{2pt}
\item \textbf{Earl (2022-09-02 00Z)}: $\sim$24~hours before Hurricane Earl's official tropical-storm genesis on September~3.
\item \textbf{Fiona (2022-09-09 12Z)}: highest observed FFS genesis rate in the dataset, within the Fiona precursor environment ($\sim$5 days before Fiona's formation).
\item \textbf{Ian 2022-09-22 00Z}: 48~hours before Hurricane Ian's tropical-storm genesis designation on 24~September.
\end{itemize}

\subsection{Feature Tracking and Tropical Contamination Screen}

The vortex tracker is a purpose-built component calibrated to the SDL-WXFormer grid.
We considered community trackers (e.g., TempestExtremes, TRACK, tobac
\citep{heikenfeld2019tobac}), but terrain artifacts and the characteristics of the
1$^\circ$ grid required customization that was more tractable to implement directly.
The warm-core screen follows the Cyclone Phase Space (CPS) formulation of
\citet{hart2003cyclone} as documented by the FSU CPS tool
(\url{https://moe.met.fsu.edu/cyclonephase/help.html}).
The screen is used mainly to remove MSLP artifacts from the terrain file and the
1$^\circ$ grid, and secondarily to remove recurving or baroclinic systems.
The tracker and screening parameters are tuned for this model and grid.

During the flux phase, the order parameter $Q$ is the global minimum MSLP over the basin domain $\Omega$, which may correspond to any co-existing tropical system. We maintain a running list of all active vortex candidates.
At each 6-hour step, Gaussian smoothing ($\sigma=1.5$ grid-points) locates the
position of each local minimum. The raw, unsmoothed MSLP value at that grid cell is the order parameter.
Existing tracks are updated by nearest-neighbor matching within $12^\circ$, with
unmatched minima below $12^\circ$N over land suppressed as terrain artifacts and
remaining unmatched minima seeding new tracks.
A track is retained for up to two consecutive timesteps of failed matching before
being dropped.
All co-existing systems are tracked simultaneously. Any storm that crosses $\lambda_0$
records a decorrelated flux crossing, regardless of whether it is the deepest system
at that moment.

Once a trajectory has been saved at $\lambda_0$ and begins the shooting phase, the
target vortex is the one that triggered the $\lambda_0$ crossing.  At each timestep,
the MSLP minimum is sought within $12^\circ$ of the vortex's last known position.
This single-vortex constraint prevents $Q$ from jumping to a stronger but unrelated
system, ensuring that the shooting trajectory measures the intensification of the
specific disturbance seeded at $\lambda_0$.

The contamination screen follows the Cyclone Phase Space framework of \citet{hart2003cyclone}
and suppresses two known sources of spurious state-$B$ arrivals at 1$^\circ$ grid spacing:
(1) terrain-induced MSLP minima from hydrostatic extrapolation errors in the imperfect
terrain file, and (2) mid-latitude baroclinic cyclones entering the domain.
The screen evaluates $-V_T^L$ and $-V_T^U$ at $\lambda_0$ crossings and when a
tracked vortex exceeds $50^\circ$N or $-10^\circ$W. Systems with $-V_T^L < 0$ or
$-V_T^U < 0$ are rejected as cold-core. In practice, a numerical tolerance permits
values down to $-100$~m before a cold-core classification is assigned, accommodating
grid noise at 1$^\circ$ grid spacing.
The rejection thresholds were calibrated by manual inspection of a subset of 2022
trajectories visually identifiable as extratropical (recurving systems and baroclinic
cyclones intruding from the northwest). They are not derived from formal CPS
validation at this grid spacing and should be treated as engineering choices rather
than physical thresholds. We err on the side of inclusion: ambiguous systems are accepted rather than discarded.

Across the full dataset, rejection rates at each interface are approximately
30\% ($\lambda_1$), 20\% ($\lambda_2$), 27\% ($\lambda_3$), and 43\% ($\lambda_4$),
with per-IC variation from 23\% to 93\% at $\lambda_4$ in the most suppressed environments.
The elevated rejection rate at $\lambda_4$ is physically plausible across the dataset: deep systems that recurve poleward often transition into extratropical cyclones while still intense, as both Earl and Fiona did in 2022.
The non-monotone pattern at earlier interfaces reflects marginal cyclones that failed
to maintain a warm core through the intensification cascade.
The high per-IC variation at $\lambda_4$ (up to 93\% in suppressed environments)
reflects that those rare vortices which deepen to 975~hPa in unfavorable environments
are disproportionately anomalous poleward-recurving tracks rather than canonical
tropical development.

\section{Results}

\subsection{Seasonal Genesis Rates and Self-Consistency Check}
\label{sec:results_val}
\begin{figure}[htbp]
\centering
\includegraphics[width=\textwidth]{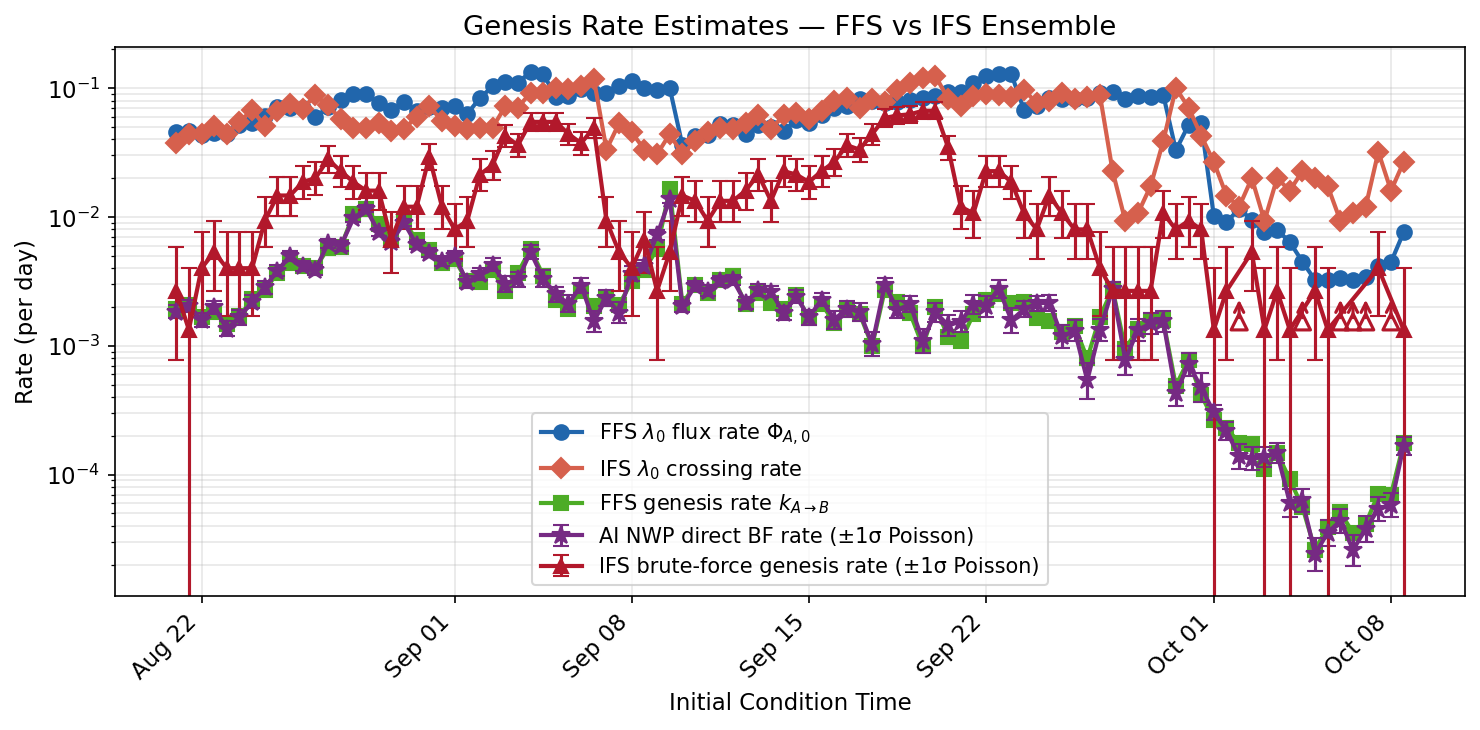}
\caption{FFS genesis rates across all 98 initial conditions (August~21 -- October~8, 2022)
on a logarithmic scale.
\textit{Blue}: $\lambda_0$ flux rate $\Phi_{A,0}$.
\textit{Orange}: IFS $\lambda_0$ crossing rate (contextual reference; not directly comparable to FFS due to different grid spacing and no CPS filter).
\textit{Green}: Full FFS genesis rate $k^{\text{FFS}}$ (flux $\times$ product of conditional probabilities).
\textit{Purple}: Direct-sampling rate $k^{\text{direct}}$ from state-$B$ arrivals during the flux phase ($\pm 1\sigma$ Poisson).
\textit{Red}: IFS brute-force genesis rate ($\pm 1\sigma$ Poisson; upward arrows denote upper bounds where zero events observed).
Agreement between green and purple (mean ratio $1.03\pm0.15$) validates the FFS method.
The FFS rate resolves a seasonal cycle spanning nearly three orders of magnitude
invisible to direct 50-member ensemble sampling.}
\label{fig:rates}
\end{figure}

Figure~\ref{fig:rates} shows FFS genesis rates across all 98 initial conditions.
The full seasonal cycle spans nearly three orders of magnitude,
from $2.6\times10^{-5}$~day$^{-1}$ during suppressed October conditions to
$1.7\times10^{-2}$~day$^{-1}$ during the peak-activity period around September~9--10
(consistent with the Fiona formation environment). Direct 50-member ensemble sampling cannot resolve this dynamic range.
For context, the ECMWF IFS 50-member ensemble \citep{ifs_ens_docs} analyzed
over the same 98 initial conditions crosses the $\lambda_0$ threshold in 25.4\% of
members on average. We include this in Figure~\ref{fig:rates} as a \textit{qualitative
sanity check} only. A direct rate comparison between FFS (SDL-WXFormer, 1$^\circ$) and IFS
(0.25$^\circ$ physics, 1.5$^\circ$ output) is not appropriate: IFS does not apply
the decorrelated-crossing requirement defining $\lambda_0$ flux, does not use a CPS
filter, and operates at different grid spacing with different physics.
The IFS figures confirm that strong MSLP deepening occurs in the model at a
plausible frequency. Quantitative rate comparisons between the two systems
are deferred to future work with matched configurations.

\begin{figure}[htbp]
\centering
\includegraphics[width=\textwidth]{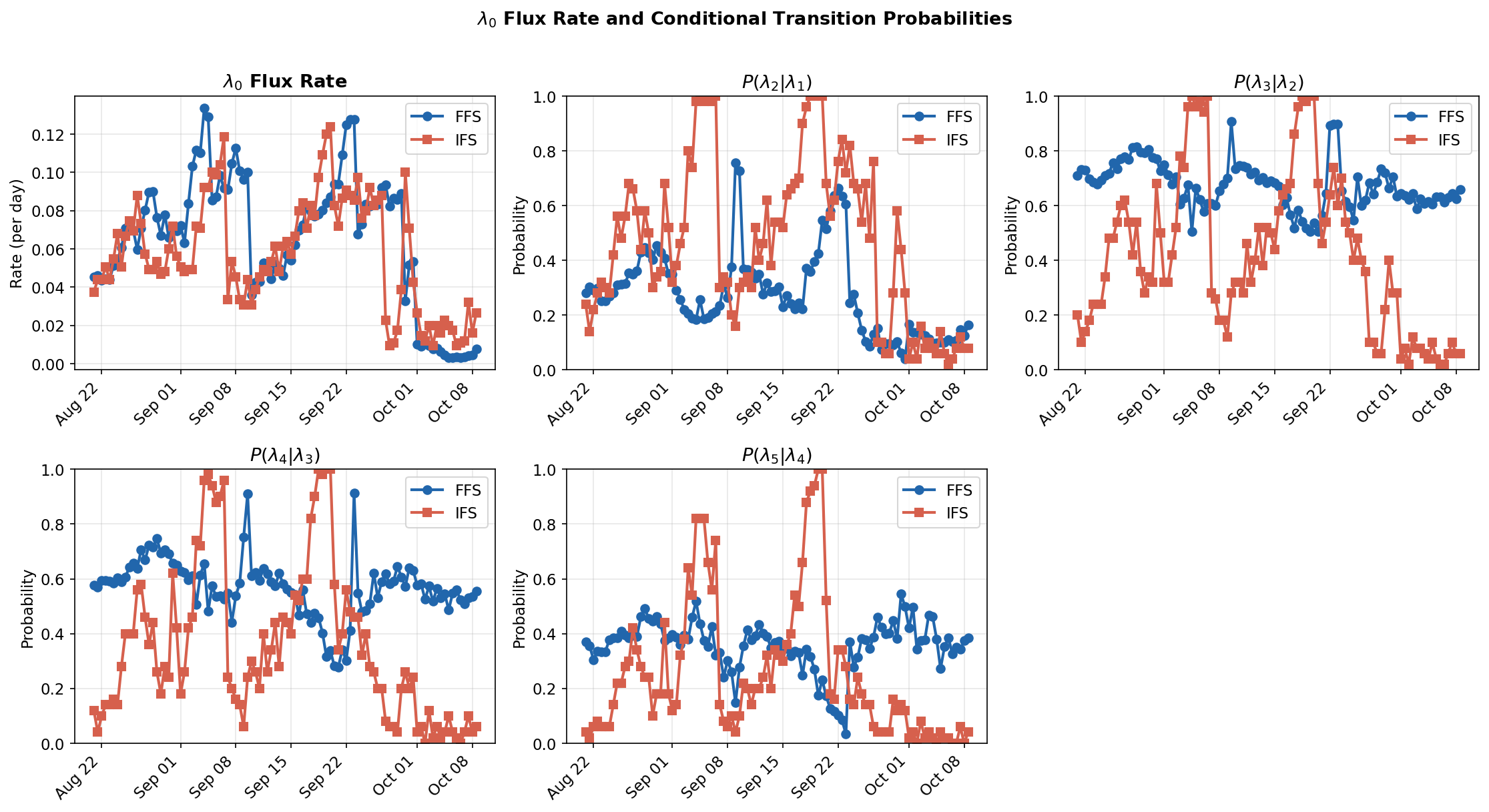}
\caption{Seasonal evolution of FFS conditional crossing probabilities $P(\lambda_i|\lambda_{i-1})$
for all 98 initial conditions (August~21 -- October~8, 2022). The first panel shows the $\lambda_0$
flux rate. Subsequent panels show each $P_i$ between successive shooting interfaces.
The dashed line marks the all-season mean. $P_1$ (initial organization, $\lambda_0\to\lambda_1$)
varies most strongly with the seasonal cycle, with a pronounced peak in early September coinciding
with the Fiona formation environment. $P_4$ (final intensification, $\lambda_3\to B$) is
suppressed across the season, limiting the overall genesis rate even in active periods.
The three case-study ICs (Earl~09-02, Fiona~09-09, Ian~09-22) are indicated by vertical markers.}
\label{fig:interface_probs}
\end{figure}

The seasonal evolution of each conditional probability $P_i$
(Fig.~\ref{fig:interface_probs}) reveals two distinct regimes within the intensification
cascade. $P_1$ (initial organization) is the most variable across ICs, with a standard
deviation of 0.154 compared to 0.086--0.107 for $P_2$--$P_3$. It spans from below 0.1
in suppressed environments to 0.726 in the peak-activity Fiona IC, reflecting the strong
sensitivity of early-stage organization to the large-scale atmospheric state. By contrast,
$P_4$ (final intensification to 975~hPa) has a season-mean of 0.356, well below $P_2$
and $P_3$, limiting the overall genesis rate even
when earlier-stage probabilities are elevated. These two interfaces represent physically
distinct bottlenecks. Which one dominates varies by environment, as the case studies below
illustrate.

\begin{figure}[htp]
\centering
\includegraphics[width=0.8\textwidth]{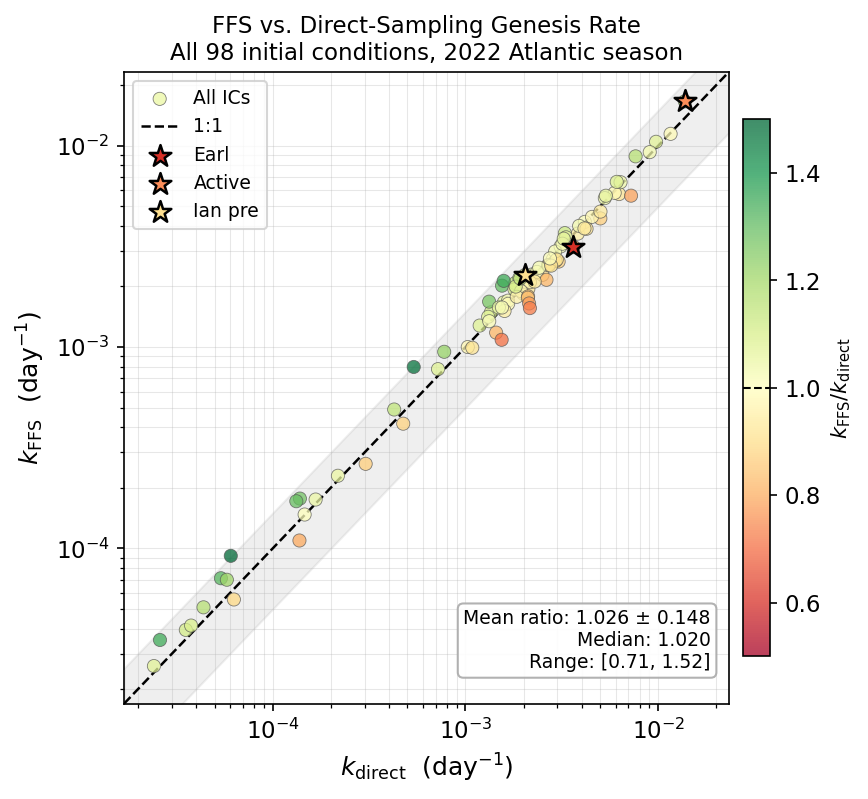}
\caption{Log-log scatter of FFS genesis rate $k^{\text{FFS}}$ versus independent
direct-sampling rate $k^{\text{direct}}$ for all 98 initial conditions.
Points are coloured by the ratio $k^{\text{FFS}}/k^{\text{direct}}$: green indicates
near-unity agreement; red/blue indicate over- or under-estimation.
The black dashed line is the 1:1 reference; shaded band spans $\pm$50\%.
Three case-study ICs are marked with stars (Earl: red; Fiona: orange; Ian: yellow).
The tight clustering along the 1:1 line across three orders of magnitude confirms
correct implementation of the conditional probability factorization.
Inset statistics: mean ratio $1.03\pm0.15$, median $1.02$, range $[0.71, 1.52]$.}
\label{fig:rate_scatter}
\end{figure}

The primary self-consistency check is the agreement between $k^{\text{FFS}}$ and the
independent direct-sampling rate $k^{\text{direct}}$ (Figure~\ref{fig:rate_scatter}).
These estimates are derived from entirely separate components of the same simulation:
$k^{\text{FFS}}$ is a product of flux and shooting probabilities, while $k^{\text{direct}}$
counts raw state-$B$ arrivals during the flux phase.
Across all 98 initial conditions, the ratio $k^{\text{FFS}}/k^{\text{direct}} = 1.03 \pm 0.15$
(mean $\pm$ std; median 1.02; range 0.71--1.52; Figure~\ref{fig:rate_scatter}).
The near-unity mean and median indicate no systematic bias in the factorization.
The standard deviation of 0.15 and the factor-of-two range reflect the Poisson
sampling noise inherent in $k^{\text{direct}}$ (which has few counts at suppressed ICs),
not a bias in $k^{\text{FFS}}$.
Critically, this agreement holds across three orders of magnitude of genesis probability,
which would be expected only if (1) the interface spacing produces well-conditioned
probability estimates, (2) the ensemble sizes are adequate, and (3) the CPS screen
does not introduce systematic bias.
We interpret this agreement as confirming the correct implementation of the conditional
probability factorization within the SDL-WXFormer framework.
It is important to note what this check does and does not establish.
It confirms that the FFS bookkeeping arithmetic is internally consistent: the product
$\Phi_{A,0}\prod P_i$ recovers the same rate that direct state-$B$ counting yields.
It does \emph{not} constitute verification against observed climatology, nor does it
validate the Markov assumption, nor does it guarantee that the absolute
rate values are unbiased with respect to the real atmosphere.
Computational enhancement factors range from $3\times$ (most active IC) to $140\times$
(most suppressed), with geometric mean of $14\times$.

\subsection{Case Studies: Distinct Genesis Bottlenecks}
\begin{figure}[htbp]
\centering
\includegraphics[width=\textwidth]{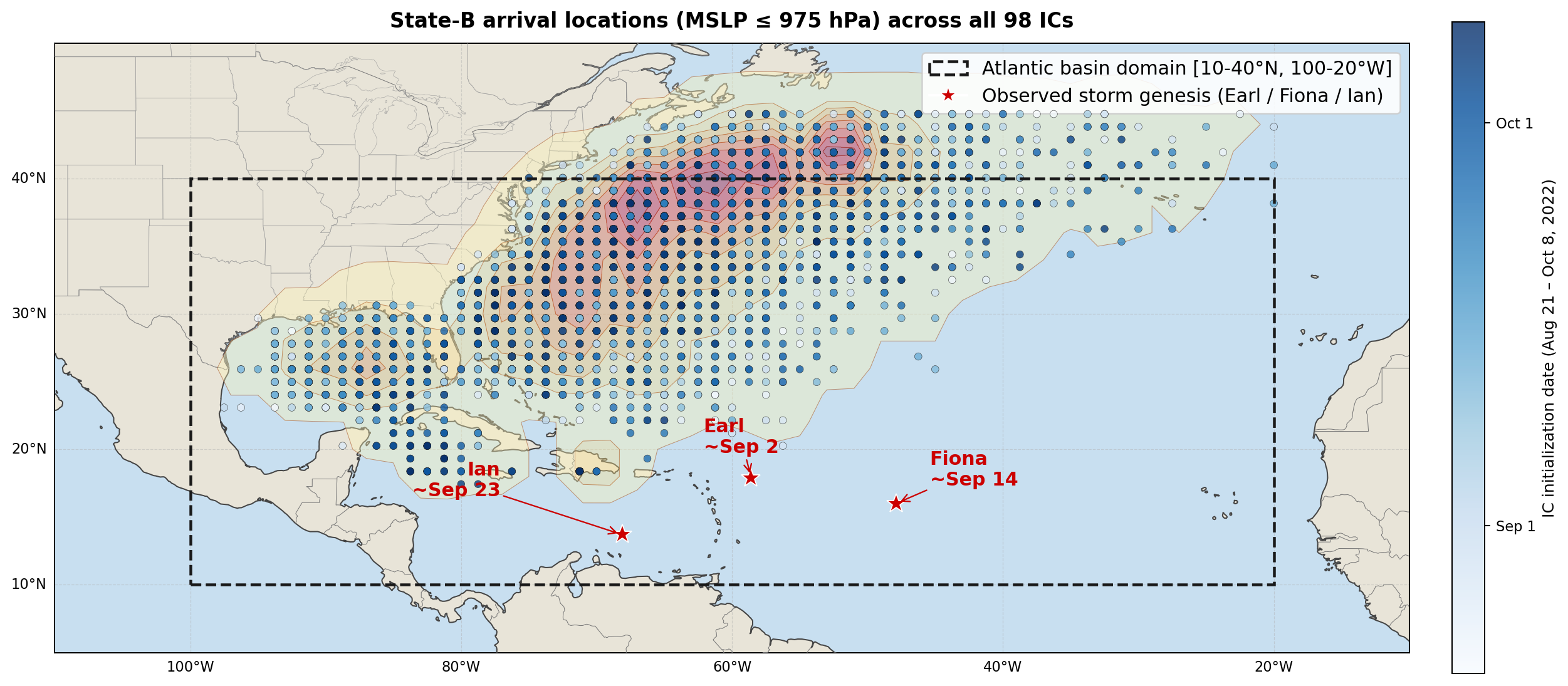}
\caption{Geographic distribution of state-$B$ arrival locations (MSLP $\leq$ 975~hPa)
across all 98 initial conditions.
Each point marks where a reactive trajectory first reached state~$B$. Colour encodes
the IC initialization date from August~21 (light) to October~8 (dark).
The 2$^\circ$ density contours (background) show the spatial clustering of genesis events
across the season. Approximate observed genesis locations for Earl, Fiona, and Ian are
annotated for reference. The distribution is concentrated in the tropical Atlantic,
Caribbean, and Gulf of Mexico, consistent with observed 2022 Atlantic storm tracks,
and shifts westward later in the season as Ian-like Gulf environments become active.}
\label{fig:stateb_geo}
\end{figure}

\begin{figure}[htbp]
\centering
\includegraphics[width=\textwidth]{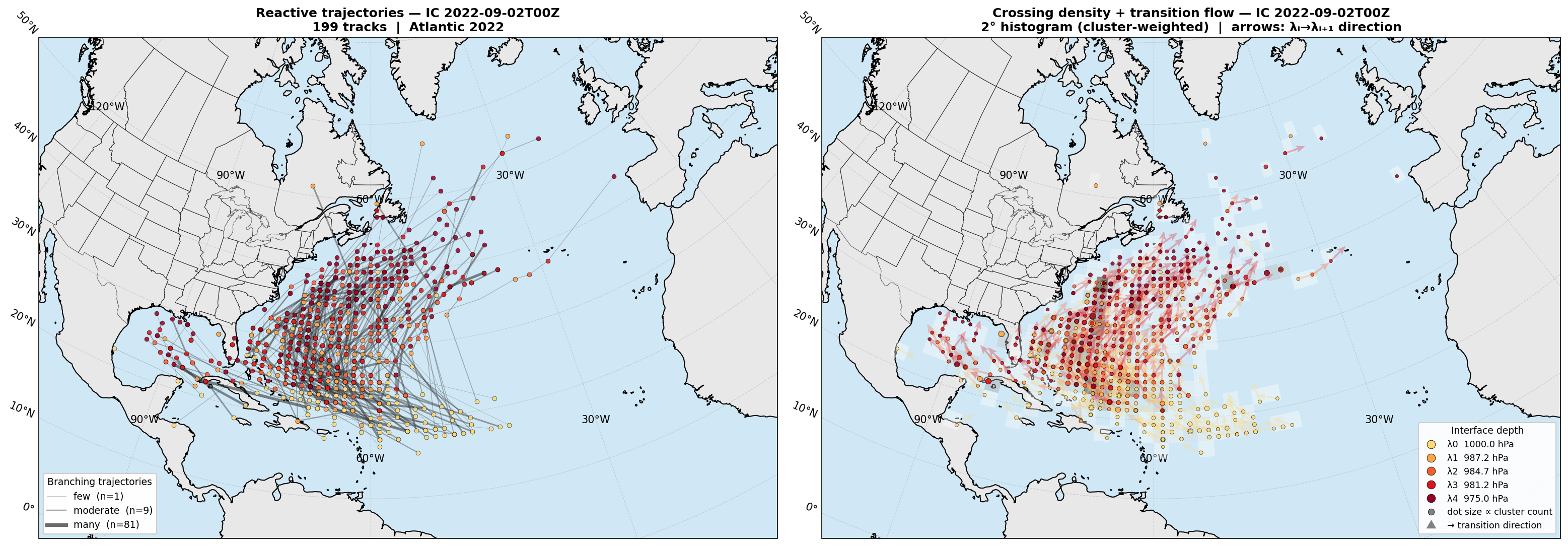}
\caption{Reactive genesis trajectories for the Earl IC (2022-09-02T00Z).
\textit{Left}: Individual tracks that cross from $\lambda_0$ through to state~$B$,
color-coded by branching factor (cluster weight). The 199 reactive pathways sample
a broad region spanning the tropical Atlantic, Caribbean, and western subtropical Atlantic.
\textit{Right}: Cluster-weighted crossing density (2$^\circ$ histogram) and mean
$\lambda_n\to\lambda_{n+1}$ transition direction (arrows), revealing the geographic regions
most active in the intensification cascade for this environment.
CPS-rejected (extratropical) trajectories are excluded from both panels.}
\label{fig:reactive_traj}
\end{figure}

\begin{table}[htbp]
\centering
\caption{FFS statistics for three case-study initial conditions.
$\Phi_{A,0}$: flux rate. $P_i$: conditional crossing probability at interface $\lambda_i$.
$k^{\text{FFS}}$: full genesis rate (Eq.~\ref{eq:ffs_rate}). $k^{\text{direct}}$:
independent direct-sampling rate. Ratio: $k^{\text{FFS}}/k^{\text{direct}}$.
Speedup: enhancement over direct sampling for equivalent 20\% rate uncertainty.
Bottom row: statistics across all 98 ICs.}
\label{tab:results}
\resizebox{\textwidth}{!}{%
\begin{tabular}{lccccccccc}
\toprule
IC & $\Phi_{A,0}$ & $P_1$ & $P_2$ & $P_3$ & $P_4$ & $k^{\text{FFS}}$ & $k^{\text{direct}}$ & Ratio & Speedup \\
   & (day$^{-1}$) & & & & & (day$^{-1}$) & (day$^{-1}$) & & \\
\midrule
Earl (09-02T00Z)   & 0.0837 & \textbf{0.258} & 0.679 & 0.596 & 0.359 & $3.13\times10^{-3}$ & $3.61\times10^{-3}$ & 0.87 & 10$\times$ \\
Fiona (09-09T12Z)  & 0.1002 & 0.726 & 0.906 & 0.910 & \textbf{0.276} & $1.66\times10^{-2}$ & $1.37\times10^{-2}$ & 1.21 &  3$\times$ \\
Ian (09-22T00Z)    & 0.1247 & 0.663 & 0.893 & \textbf{0.302} & \textbf{0.102} & $2.28\times10^{-3}$ & $2.04\times10^{-3}$ & 1.12 & 19$\times$ \\
\midrule
All-98 mean$\pm$SD  & $0.065\pm0.034$ & $0.278\pm0.154$ & $0.674\pm0.086$ & $0.569\pm0.107$ & $0.356\pm0.093$ & $2.6\times10^{-3}$ & $2.5\times10^{-3}$ & $1.03\pm0.15$ & $14\times$$^{\dagger}$ \\
\bottomrule
\multicolumn{10}{l}{$^{\dagger}$Geometric mean; range 3--140$\times$. Speedup distribution is right-skewed; geometric mean is the appropriate summary.}
\end{tabular}%
}
\end{table}
Table~\ref{tab:results} presents FFS statistics for the three case-study initial
conditions. The most striking feature is the variable position of the rate-limiting
step, encoded in which conditional probability is smallest.
Figure~\ref{fig:stateb_geo} shows the geographic distribution of state-$B$ arrivals
across all 98 ICs, confirming that the model's genesis events cluster in physically
plausible locations (tropical Atlantic, Caribbean, Gulf of Mexico) and shift
geographically with the seasonal cycle.
For the Earl IC, Fig.~\ref{fig:reactive_traj} shows the 199 reactive
pathways that successfully reached state~$B$, illustrating the geographic diversity
of intensification trajectories sampled by FFS from a single background state.

For the \textbf{Earl IC}, $P_1 = 0.258$ is the dominant bottleneck: roughly 1 in 4
trajectories reaching 1000~hPa continue to organize through to 987~hPa in this environment. This is
physically consistent with Earl's pre-genesis state $\sim$24~hours before classification.
At that point the system had established a coherent low-level circulation but had not yet
achieved the sustained convective organization needed to cross the first intensification
threshold.

For the \textbf{Fiona IC} (September~9), all four conditional probabilities
are elevated ($\geq0.27$), and the product $\prod P_i = 0.166$ is $\sim$4--9$\times$
larger than the other two cases. The dominant bottleneck shifts to $P_4 = 0.276$
(final intensification to 975~hPa), reflecting an environment that readily organizes
disturbances but where crossing the final intensification threshold remains uncertain.
The FFS rate ($1.66\times10^{-2}$~day$^{-1}$) is high enough that direct sampling
is also feasible (speedup $3\times$), providing a strong cross-validation: both methods
agree to within 21\% for this most-active IC.

The \textbf{Ian IC} presents the most physically interesting pattern.
$P_1 = 0.663$ and $P_2 = 0.893$ indicate that the environment readily supports initial
organization through 985~hPa, consistent with the warm, moist, low-shear conditions
present over the Caribbean and western Atlantic on September~22. But $P_3 = 0.302$
and $P_4 = 0.102$ reveal a hard barrier in the final intensification steps (985~hPa
to 975~hPa). This is physically consistent with Ian's trajectory: the storm was still
organizing over the Caribbean on September~22--23, entering the Gulf of Mexico's warm
core environment only after September~25, when conditions supported rapid intensification.
The FFS framework identifies the final-stage intensification barrier \textit{two days before}
Ian's tropical-storm designation, a diagnostic that direct ensemble sampling cannot resolve.

\subsection{FFS Shooting Trees}

\begin{figure}[htbp]
\centering
\includegraphics[width=\textwidth]{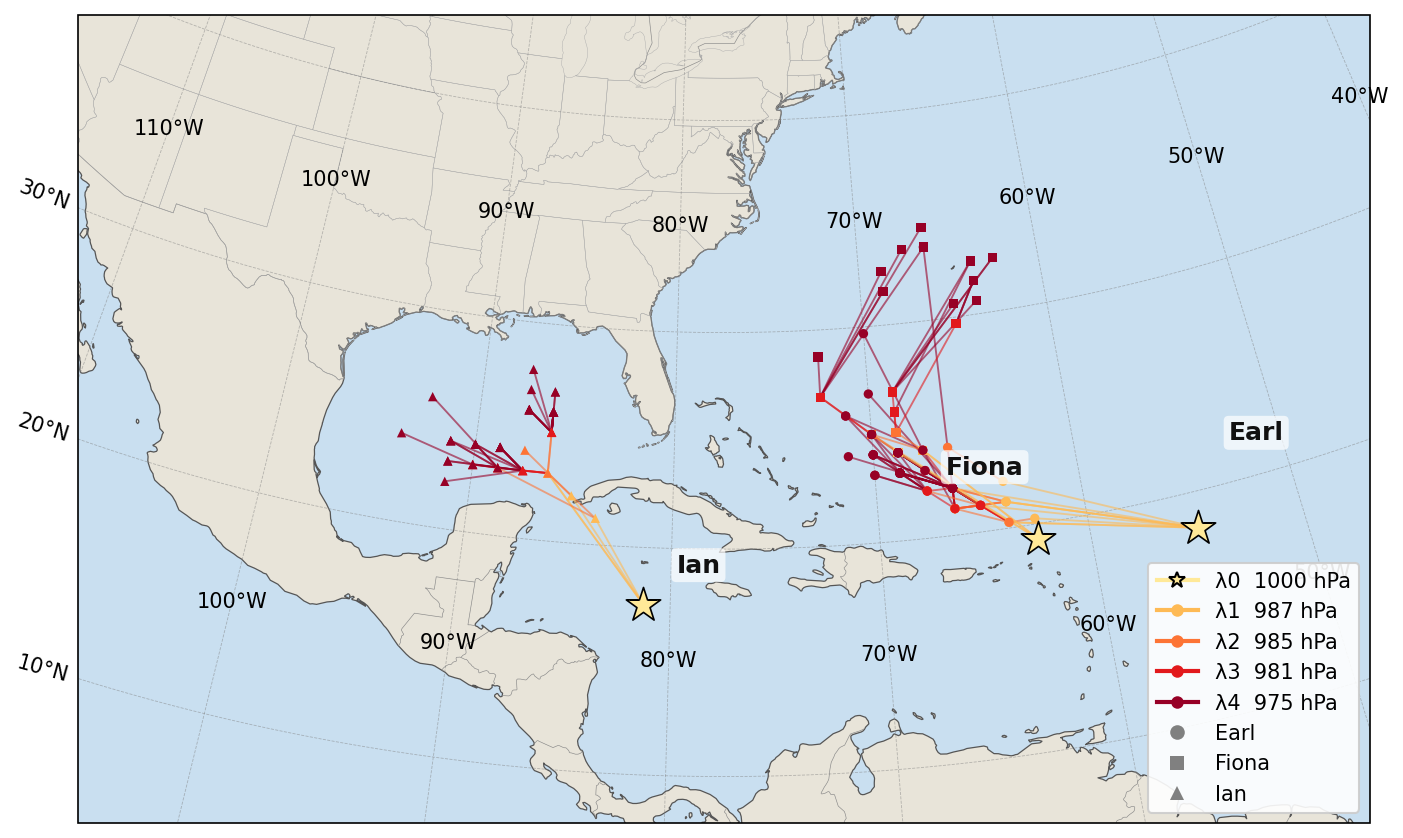}
\caption{FFS shooting trees for the three case-study ICs on a single combined map.
Each star ($\lambda_0$ seed) and its descendant branches are plotted together.
Branch colour encodes interface level from $\lambda_1$ (yellow-orange) to $\lambda_4$
(dark red), using the same colour scale for all three ICs.
IC labels identify each seed:
Earl (2022-09-02T00Z, seed near 18$^\circ$N in the eastern Atlantic, 24~hours
before Earl's genesis),
Fiona (2022-09-09T12Z, seed near 19$^\circ$N in the central Atlantic),
and Ian (2022-09-22T00Z, seed near 17$^\circ$N in the western Caribbean).
Earl trajectories fan westward consistent with Earl's observed west-northwest track;
Fiona trajectories track westward into the Caribbean consistent with Fiona's
subsequent path; Ian trajectories fan into the Gulf of Mexico in two clusters:
a west-track group toward the Bay of Campeche and a north-track group toward Florida
(see Fig.~\ref{fig:ian_gulf_composites}).}
\label{fig:ffs_trees}
\end{figure}

Figure~\ref{fig:ffs_trees} shows representative FFS shooting trees for the three
case-study ICs. The point is not the exact geometry of any one branch. The point is
that one $\lambda_0$ crossing can lead to many distinct downstream pathways, and FFS
samples that branching explicitly.

Importantly, the IC date labels are \textit{initialization} times, not genesis dates.
Each FFS simulation runs 15 days forward from its IC. Each node in the tree represents
the storm feature at a different future lead time within that window.
A single seed at $\lambda_0$ therefore spawns descendants that span a range of calendar
dates, geographic positions, and atmospheric states, all sampled from one background IC.

All three trees are displayed on a single combined map. Branch colours encode
interface level ($\lambda_1$ through $\lambda_4$) using a shared yellow-orange-red
scale, and each IC is identified by a text label.
The Earl tree (2022-09-02T00Z, 24~hours before Earl's genesis) seeds near 18$^\circ$N
in the eastern Atlantic and fans westward, consistent with Earl's observed
west-northwest track toward the Leeward Islands.
The Fiona tree (2022-09-09T12Z, the peak-activity IC within the Fiona precursor
environment) seeds near 19$^\circ$N in the central Atlantic and fans westward
into the Caribbean, consistent with Fiona's observed track.
The Ian tree (2022-09-22T00Z, 48~hours before Ian's NHC tropical-storm designation)
seeds near 17$^\circ$N in the western Caribbean and fans into the Gulf of Mexico
in two distinct track clusters: a west-track group terminating near the Bay of
Campeche and a north-track group tracking northward toward Florida.

\begin{figure}[htbp]
\centering
\includegraphics[width=\textwidth]{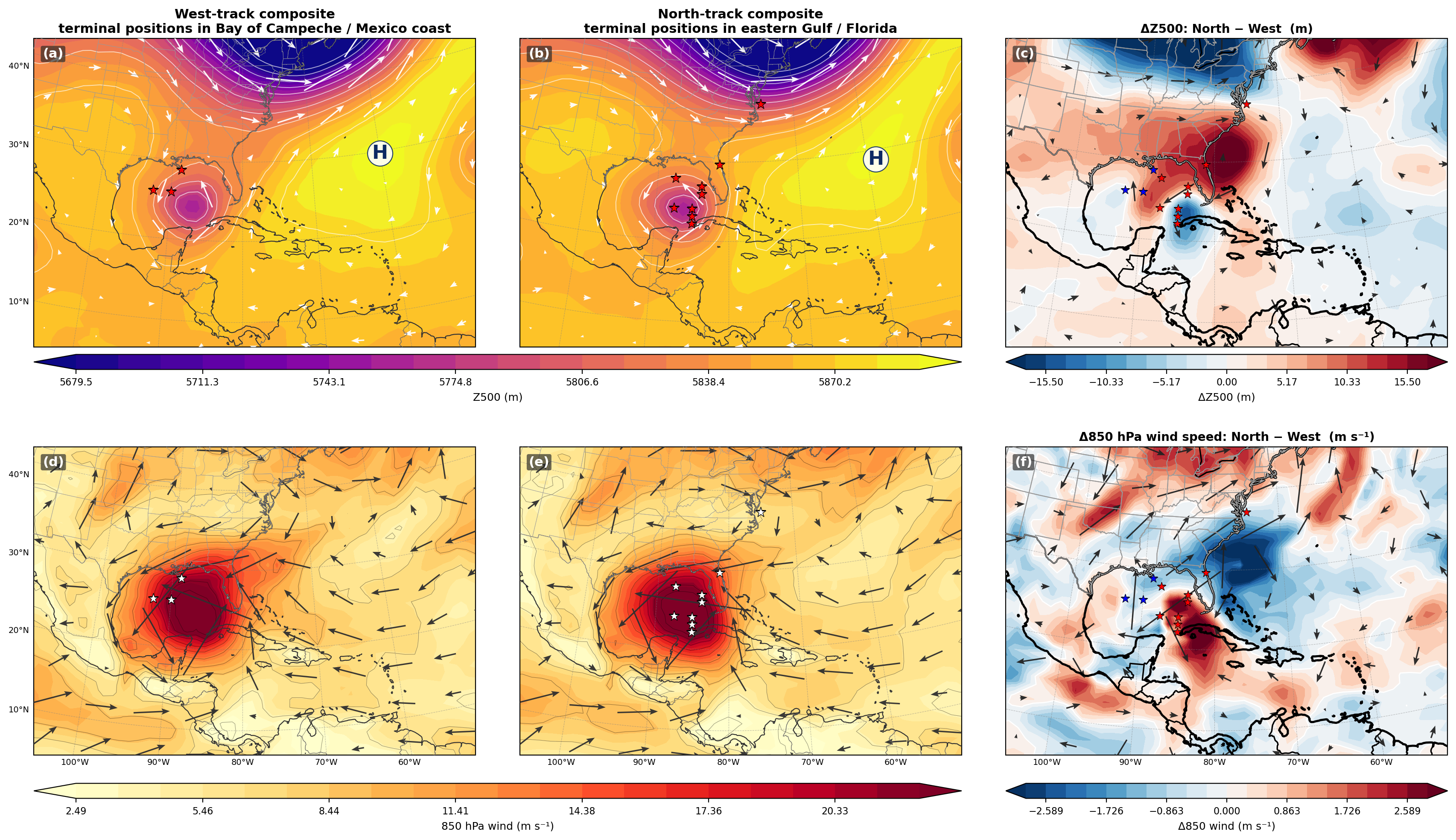}
\caption{Composite 500~hPa geopotential height (filled, row 1) with 500~hPa wind
vectors and 850~hPa wind speed (filled, row 2) with 850~hPa wind vectors, evaluated
at the $\lambda_2$ interface (984.7~hPa) for Ian IC (2022-09-22T00Z) FFS trajectories
split by terminal track. Left column: west-track trajectories (terminal longitude west
of $88^\circ$W; Bay of Campeche or Mexican coast). Center column: north-track
trajectories (terminal longitude between $88^\circ$W and $75^\circ$W; eastern Gulf
or Florida coastline). Right column: pointwise difference (north $-$ west). The
circled \textbf{H} marks the ridge center in each absolute composite. Red stars show
state-$B$ terminal positions; blue (red) stars in the difference panels mark west-track
(north-track) terminals. The north-track composite places the mid-tropospheric ridge
further northeast and with greater amplitude, producing the stronger anticyclonic flow
that steers Ian northward.}
\label{fig:ian_gulf_composites}
\end{figure}

Ian's tree branches into two geographically distinct track groups at the $\lambda_2$
interface (984.7~hPa). West-track trajectories terminate in the Bay of Campeche or
along the Mexican coast (final longitude west of $88^\circ$W), while north-track
trajectories terminate in the eastern Gulf or along the Florida coastline (final
longitude between $88^\circ$W and $75^\circ$W). Figure~\ref{fig:ian_gulf_composites}
composites the 500~hPa geopotential height and 850~hPa steering flow at $\lambda_2$
separately for each group. In the north-track composite the mid-tropospheric ridge is
displaced further northeast and is more intense, producing stronger anticyclonic flow
that deflects Ian's steering current northward into the warm Gulf of Mexico environment.
In the west-track composite the ridge sits more centrally over the Gulf, allowing a
more zonal flow to carry Ian westward. The contrast resembles, but does not meet the
criteria for, omega blocking: both composites show an amplified Atlantic high that
compresses the steering channel, increasing sensitivity to the ridge position at
$\lambda_2$.

\subsection{Committor Curves and Spatial Genesis Fields}

\begin{figure}[htbp]
\centering
\includegraphics[width=0.85\textwidth]{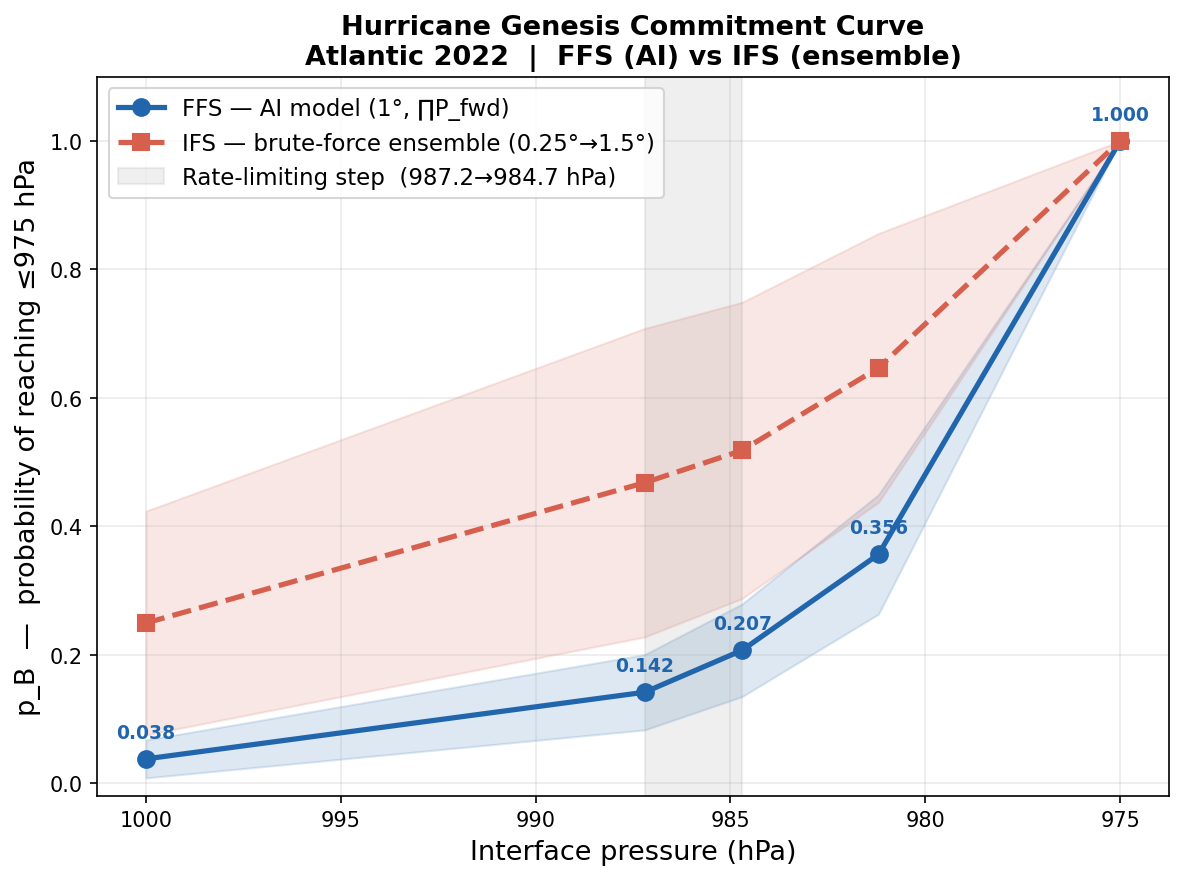}
\caption{Genesis committor curve $p_B(\lambda_i)$ — the probability of eventually reaching
state~$B$ ($\leq$975~hPa) given arrival at interface $\lambda_i$ — averaged across all 98
initial conditions.
\textit{Blue}: FFS AI model mean $\pm 1\sigma$ across ICs.
\textit{Orange}: IFS brute-force ensemble mean $\pm 1\sigma$ across ICs.
The shaded bands show IC-to-IC variability. The annotated values are the all-season FFS means.
The grey band highlights the rate-limiting step (largest single drop), which falls across
the first shooting step ($\lambda_0\to\lambda_1$) for the full 2022 Atlantic season.
The FFS and IFS curves agree in the overall shape while differing in magnitude,
consistent with the grid spacing and physics differences between the two systems.}
\label{fig:committor}
\end{figure}

Figure~\ref{fig:committor} shows the committor curve $p_B(\lambda_i)$, the probability
of eventually reaching state~$B$ given arrival at interface $\lambda_i$
\citep{bolhuis2002transition, rotskoff2018machine}, averaged across all 98 initial conditions for both the
FFS (AI model) and IFS (brute-force ensemble).
The all-season FFS curve shows a near-linear decline with the steepest drop across
the first shooting step ($\lambda_0\to\lambda_1$), consistent with the
seasonal mean of $P_1$ being the lowest of the four conditional probabilities
(Table~\ref{tab:results}).
The FFS and IFS curves agree in overall \textit{shape}: both show the steepest committor
drop at the same interface level, and both identify the same rate-limiting stage of the
intensification cascade. This suggests that the bottleneck is not unique to
SDL-WXFormer.
Note that the comparison is complicated by the different ensemble generation mechanisms
(learned stochastic layers vs.\ singular-vector perturbations) and the 1$^\circ$ vs.\ 1.5$^\circ$
grid spacing asymmetry, so quantitative agreement is not expected and not claimed.
The FFS committors are systematically higher in \textit{magnitude}, consistent with
SDL-WXFormer's higher absolute genesis rate at 1$^\circ$ relative to IFS archived at 1.5$^\circ$
(the coarser IFS output smooths MSLP minima, effectively raising the intensification barrier).

\subsection{Physics Along a Reactive Pathway}

\begin{figure}[htbp]
\centering
\includegraphics[width=\textwidth]{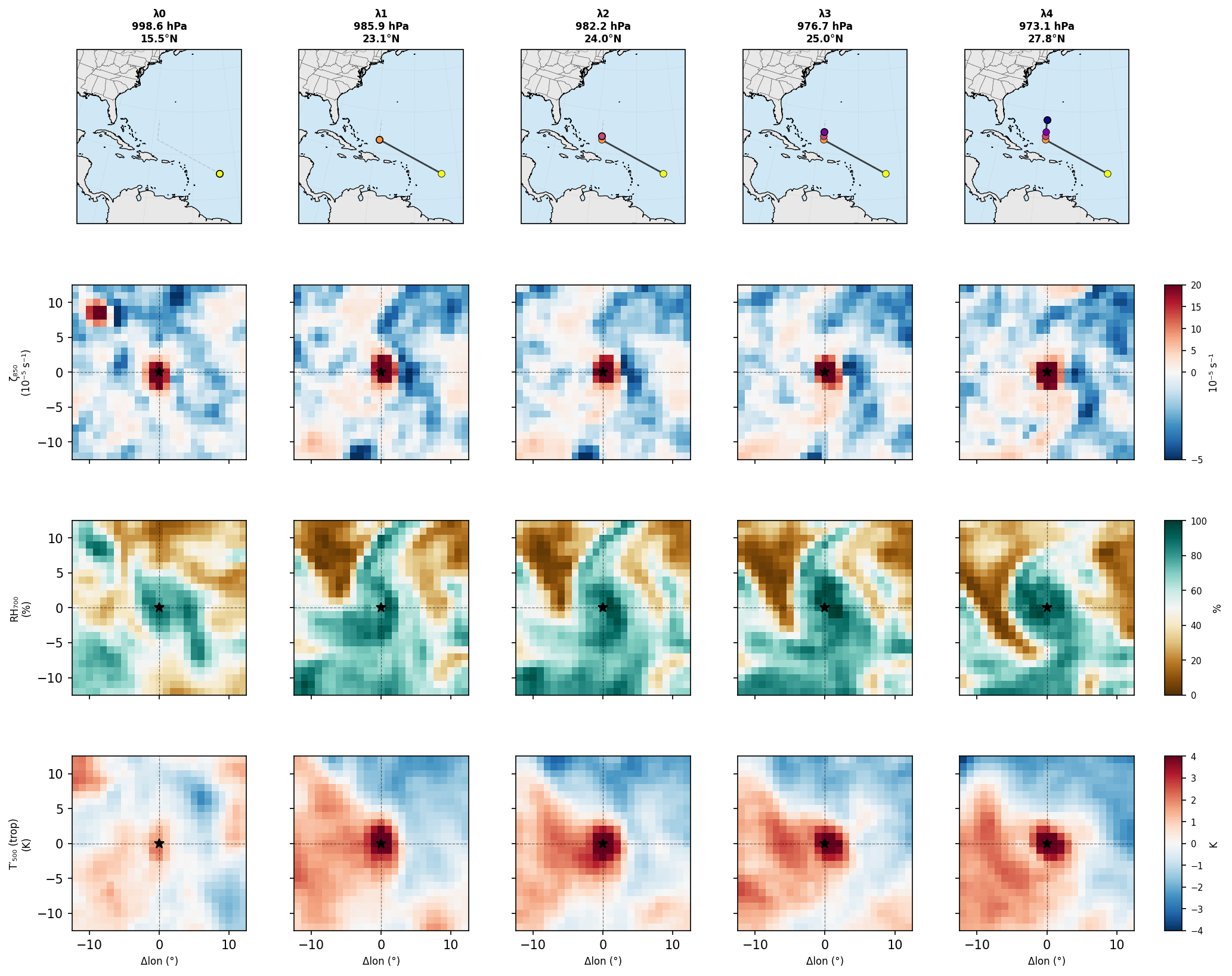}
\caption{Storm-centred physics along a representative reactive pathway for the Earl IC
(2022-09-02T00Z, pathway~1079), $\lambda_0$ through $\lambda_4$.
Each column shows one interface crossing. Column headers give the MSLP and storm latitude.
The vortex originates near 16$^\circ$N in the eastern Atlantic at $\lambda_0$ and tracks
west-northwestward to 28$^\circ$N at $\lambda_4$, consistent with Earl's observed
tropical Atlantic genesis and subsequent westward motion.
\textit{Row 1}: 850-hPa relative vorticity $\zeta_{850}$ ($10^{-5}$~s$^{-1}$).
\textit{Row 2}: 700-hPa relative humidity RH$_{700}$ (\%).
\textit{Row 3}: 500-hPa warm-core anomaly $T'_{500}$ (K, relative to 0--30$^\circ$N mean).
The persistent cyclonic vorticity and warm-core anomaly at each interface confirm continuous
vortex identity and warm-core structure through state~$B$.
Pathways where the warm core collapses at high latitudes are screened out by the CPS filter
and not included in the genesis rate calculation.}
\label{fig:physics}
\end{figure}

Figure~\ref{fig:physics} traces the storm-centred thermodynamic and kinematic evolution
along a representative Earl reactive pathway (pathway~1079) from $\lambda_0$ through state~$B$.
The vortex originates near 16$^\circ$N in the eastern Atlantic and tracks west-northwestward
to 28$^\circ$N as it intensifies from $\sim$999~hPa to 973~hPa, consistent with Earl's
observed tropical Atlantic genesis and subsequent westward motion.
The persistent cyclonic vorticity ($\zeta_{850}$) and warm-core anomaly ($T'_{500}$) at
each interface confirm continuous vortex identity and warm-core structure through state~$B$.
These storm-centred panels reveal the physical mechanisms operating at each stage,
connecting the abstract conditional probabilities in Table~\ref{tab:results} to concrete
dynamical processes.

\subsection{Storm-Centred Physics Composites}

\begin{figure}[htbp]
\centering
\includegraphics[width=\textwidth]{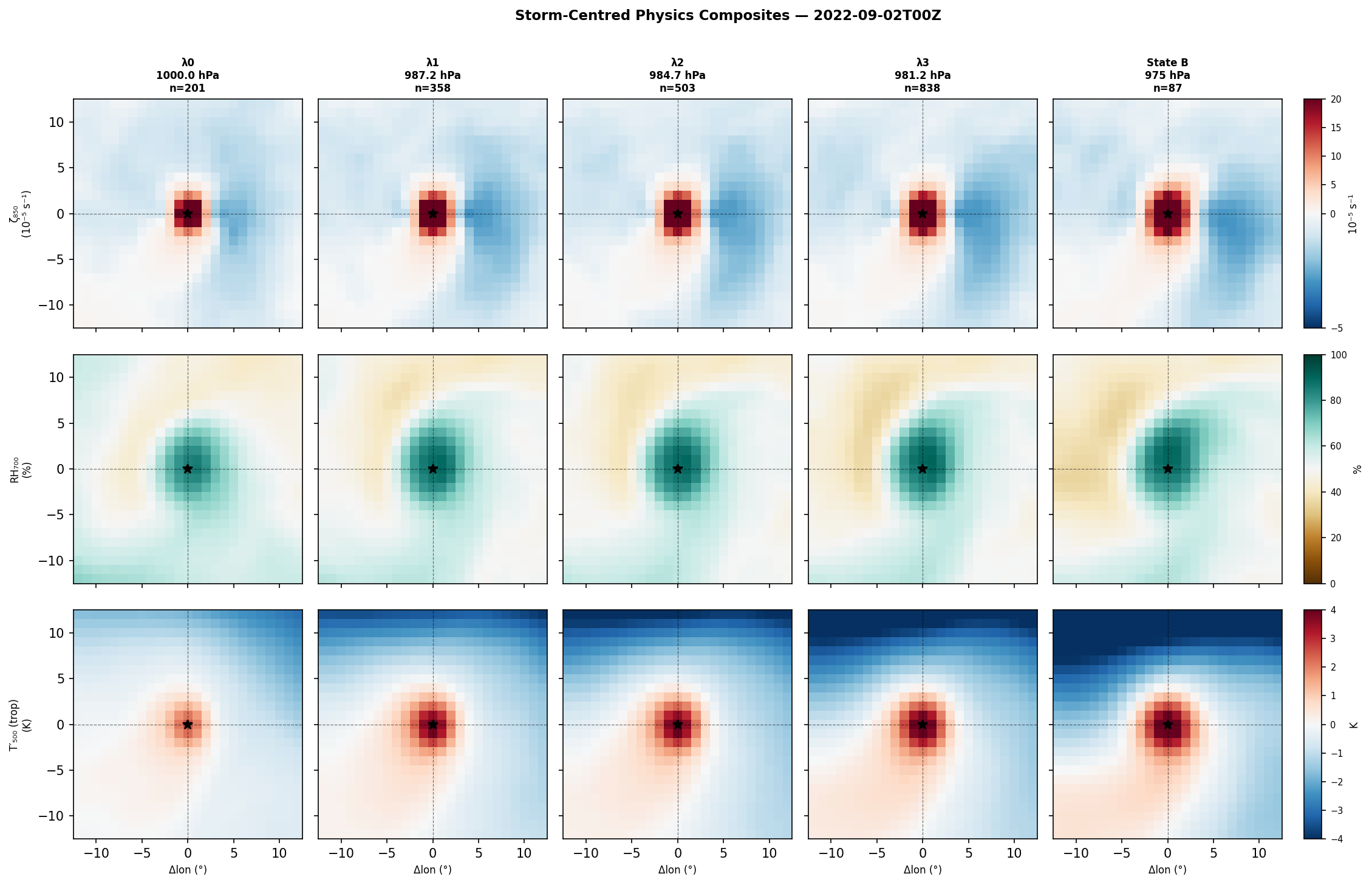}
\caption{Storm-centred physics composites along reactive pathways for the Earl IC
(2022-09-02T00Z). Columns show interfaces $\lambda_0$ through state~$B$. Sample sizes
$n$ in each column header. Rows show: (1) 850-hPa relative vorticity $\zeta_{850}$
($10^{-5}$~s$^{-1}$); (2) 700-hPa relative humidity RH$_{700}$ (\%);
(3) 500-hPa warm-core anomaly $T'_{500}$ (K).
The weaker vorticity and warm-core organisation at $\lambda_1$ reflects Earl's
lower $P_1$ value and identifies initial organisation as the rate-limiting bottleneck.}
\label{fig:composites_earl}
\end{figure}

\begin{figure}[htbp]
\centering
\includegraphics[width=\textwidth]{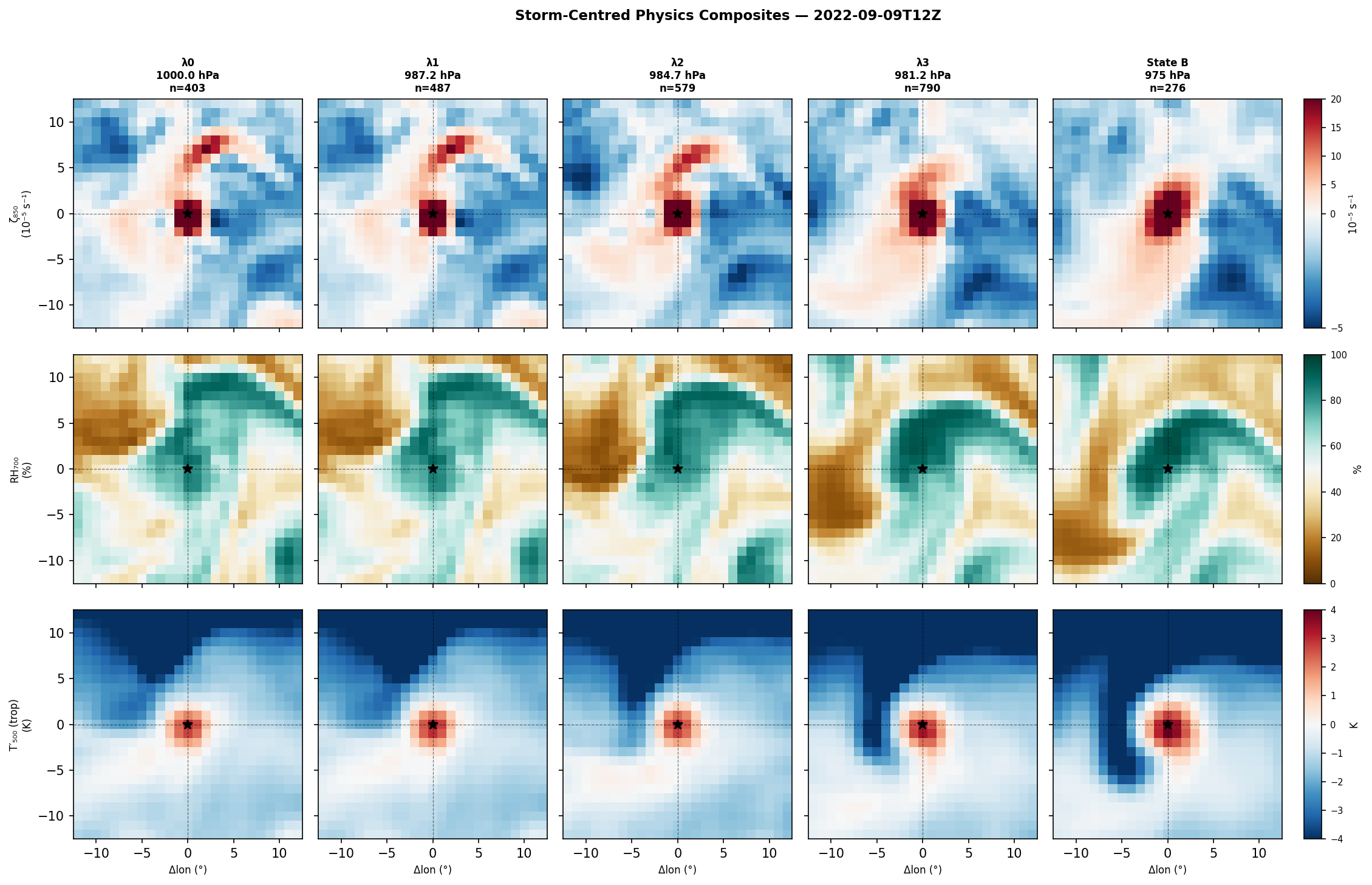}
\caption{As in Fig.~\ref{fig:composites_earl} but for the Fiona IC
(2022-09-09T12Z). Stronger vorticity and warm-core development at early interfaces
compared to Earl reflects the more favourable large-scale environment and higher $P_1$.}
\label{fig:composites_fiona}
\end{figure}

\begin{figure}[htbp]
\centering
\includegraphics[width=\textwidth]{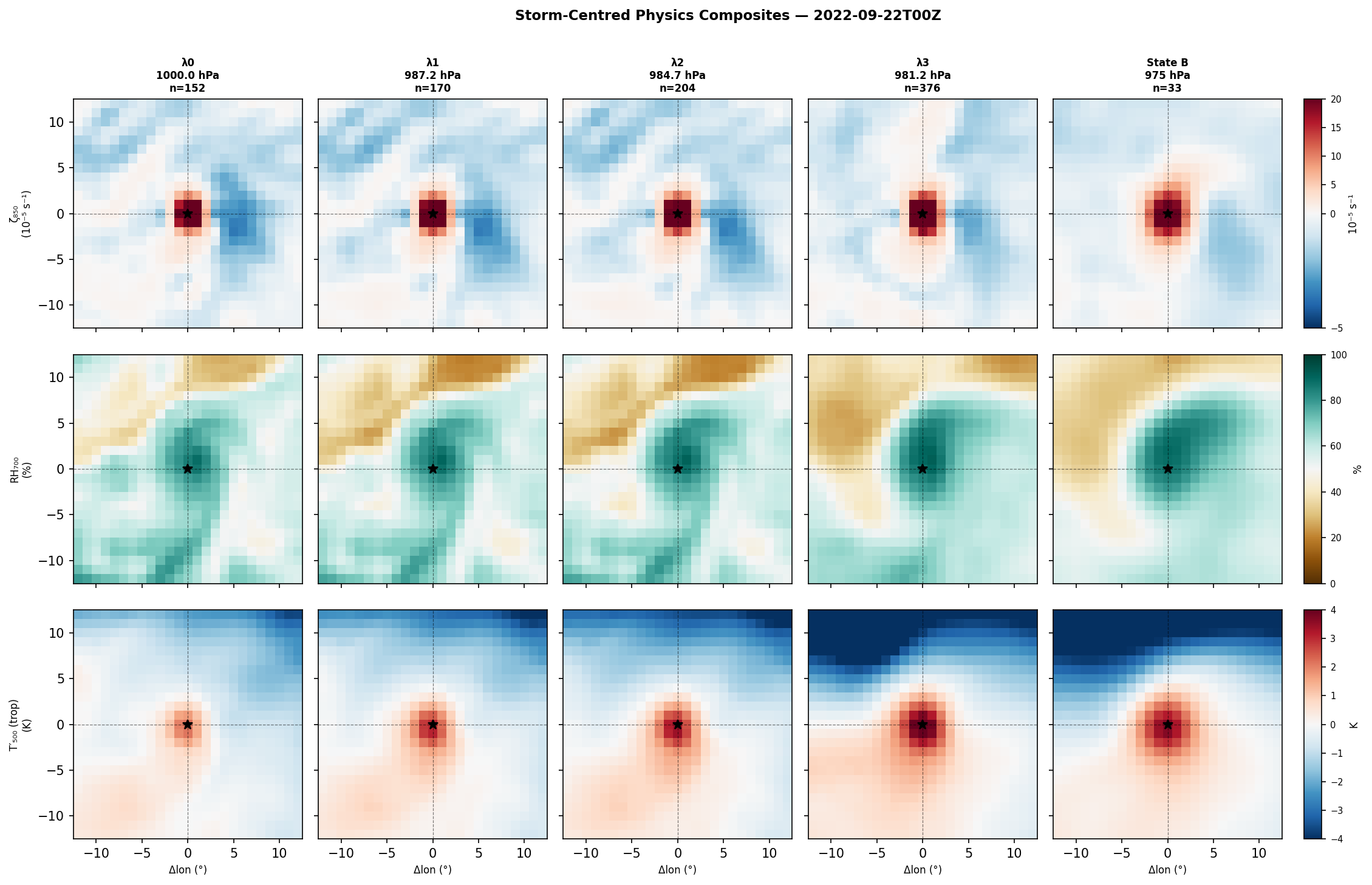}
\caption{As in Fig.~\ref{fig:composites_earl} but for the Ian IC
(2022-09-22T00Z). The $T'_{500}$ amplitude at state~$B$ is comparable across all three
ICs despite their differing genesis rates, suggesting a physically similar final
intensification pathway once the vortex reaches the near-hurricane threshold.
This composite pools all Ian reactive pathways regardless of terminal track; the
large-scale steering environment distinguishing west-track from north-track trajectories
at $\lambda_2$ is shown in Fig.~\ref{fig:ian_gulf_composites}.}
\label{fig:composites_ian}
\end{figure}

Figures~\ref{fig:composites_earl}--\ref{fig:composites_ian} show storm-centred composites
of vorticity, relative humidity, and warm-core anomaly, composited across all reactive
pathways for each case-study IC.
By averaging over all configurations that successfully reach state~$B$, these composites
reveal the mean thermodynamic and kinematic environment that supports genesis, smoothing
out pathway-to-pathway variability to expose the underlying physical signal.

All three ICs share a common structural signature: a compact positive vorticity core at
850~hPa, a saturated inner core (RH$_{700} \gtrsim 70\%$), and a growing warm-core anomaly
at 500~hPa that strengthens progressively from $\lambda_0$ through state~$B$.
The amplitude of $T'_{500}$ at state~$B$ is comparable across the three cases despite their
differing genesis rates, suggesting that once a vortex reaches state~$B$ it has traversed
a physically similar intensification pathway regardless of the environmental background.
The primary inter-IC difference appears at early interfaces. The Earl IC composites
show weaker vorticity and warm-core organisation at $\lambda_1$--$\lambda_2$ compared
to the Fiona and Ian composites, directly reflecting Earl's lower $P_1$ value
(Table~\ref{tab:results}) and confirming that the conditional probability bottleneck
has a physically interpretable composite signature.

\section{Discussion and Conclusions}

\subsection{Methodological Validation and Practical Considerations}

The central self-consistency result is that FFS and direct-sampling rates agree with
a mean ratio of $1.03\pm0.15$ across 98 initial conditions, confirming the correct
implementation of the conditional probability factorization. These estimates use separate components of
the same simulation: the FFS rate uses only the interface crossing counts
and shooting probabilities, while the direct rate uses only the raw state-$B$ arrival
counts from flux trajectories.
Their close agreement across three orders of magnitude of genesis probability confirms
that (1) the interface spacing produces well-conditioned conditional probability estimates,
(2) the ensemble sizes ($N=2000$ per interface) are sufficient (see Appendix~\ref{sec:convergence}), and (3) the CPS
contamination screen does not introduce systematic bias into the factorization.
As discussed in Section~\ref{sec:results_val}, this self-consistency check does not
constitute external validation against observed climatology, nor does it test the
Markov assumption. It confirms that the FFS bookkeeping arithmetic is implemented
correctly within the SDL-WXFormer framework.

Enhancement factors (3--140$\times$, geometric mean 14$\times$)
scale inversely with genesis probability, as expected from FFS theory \citep{allen2009forward}.
The distribution is right-skewed, so we report the geometric mean.
For the most active environments where direct sampling is feasible (speedup $3\times$), the two
methods serve as mutual validation. For suppressed environments (speedup $\sim$100$\times$),
FFS is the only practical route to a quantitative genesis rate estimate.

The bottleneck analysis says more than the rate alone. For the Earl IC,
the rate-limiting step at $\lambda_1$ (initial organization) indicates that the
environment is resistant to producing organized convection but that once organized,
the system is likely to reach hurricane strength. For the Fiona IC, the uniformly elevated probabilities indicate an environment that
readily supports each successive intensification step. The terminal barrier at
$P_4 = 0.276$ reflects that crossing the final 975~hPa threshold requires a specific
convective alignment not guaranteed even in this highly active environment.
For the Ian IC, easy initial organization but hard final intensification ($P_4 = 0.102$)
indicates that the large-scale structure favors genesis attempts, but the thermodynamic
pathway to full hurricane development requires specific mesoscale alignment not yet
present 48~hours before NHC tropical-storm designation.
These MSLP-based bottleneck identifications are corroborated by independent
thermodynamic and kinematic evidence. The storm-centred composites
(Figs~\ref{fig:composites_earl}--\ref{fig:composites_ian}) show that the interface
at which each IC's $P_i$ is smallest corresponds to a measurably weaker vorticity
core, reduced mid-level relative humidity, and less developed warm-core anomaly in
the ensemble-mean composite. This confirms that the pressure-based bottleneck
reflects a genuine physical barrier in the intensification cascade.

The CPS filter warrants clarification. At 1$^\circ$ grid spacing, formal CPS parameters
are not quantitatively validated against higher-resolution analyses, and our implementation
omits the thermal asymmetry ($B$) parameter. We use it as a contamination screen, not
as a definitive warm-core classifier. It rejects clearly extratropical systems and keeps
ambiguous ones.

\subsection{Limitations}

This study analyzes a single hurricane season (2022). Interannual variability driven
by ENSO, the AMO, and Saharan dust outbreaks is not represented. The SDL-WXFormer was
trained on ERA5 (1979--2021) at 1$^\circ$ grid spacing, which smooths intense vortex
cores and underrepresents Category~3+ hurricanes. State~$B$ is accordingly defined at
975~hPa rather than the intensities typical of major hurricane activity. Validation
against physics-based models (WRF, IFS) at higher resolution is an important next step.
Recent work confirms more broadly that current AI weather models underestimate the frequency
and intensity of record-breaking extremes relative to physics-based systems \citep{zhang2026physics},
reinforcing the value of such benchmarking.

Additionally, MSLP as an order parameter has known limitations at 1$^\circ$ grid spacing:
terrain-induced pressure artifacts (present in this model due to topographic interpolation
differences) require the CPS filter and geographic masking described in Section~2.3.
MSLP is a downstream integrator of the thermodynamic and kinematic preconditions for
genesis (warm-core development, vorticity alignment, mid-level moistening, shear
suppression) rather than a direct measure of those precursors.
As a result, the FFS rate strictly measures the frequency of organised MSLP deepening,
and the per-interface crossing probabilities reflect barriers to pressure deepening
specifically.
That said, the storm-centred composite figures (Figs~\ref{fig:composites_earl}--\ref{fig:composites_ian})
show that the MSLP-defined interfaces correspond to coherent and physically interpretable
thermodynamic states. Vorticity, mid-level relative humidity, and warm-core anomaly all
evolve progressively from $\lambda_0$ through state~$B$, confirming that the MSLP
interfaces track genuine tropical development rather than spurious pressure minima.
Nonetheless, reaction coordinates that incorporate vorticity, saturation fraction, or a
composite genesis potential index \citep[e.g.,][]{bister1998dissipative,camargo2007use}
are expected to sharpen the physical interpretability of per-interface bottleneck
diagnostics and are a natural extension for future work.

\subsection{Conclusions}

Forward Flux Sampling resolves Atlantic tropical cyclogenesis rates spanning three orders
of magnitude from 98 atmospheric initial conditions in a fast neural weather emulator.
A self-consistency check shows a mean ratio of FFS to direct-sampling rates of
$1.03\pm0.15$ across all 98 ICs, confirming the conditional probability factorization
is implemented correctly.
Computational enhancement of 3--140$\times$ (geometric mean $14\times$) makes FFS
practical across the full range of synoptic environments. The main physical result is
that the rate-limiting step in genesis varies by regime: initial tropical organization
($P_1$ bottleneck, as in Earl), uniformly elevated crossing probabilities with a terminal
intensification barrier ($P_4$ bottleneck, as in the Fiona IC), or a compound
barrier at the final intensification stages ($P_3$--$P_4$, as in the Ian IC).
These distinct bottleneck signatures encode mechanistic information about genesis pathway
dynamics that is invisible to scalar rate estimates from direct sampling.
A cross-model comparison of genesis committor curves between SDL-WXFormer and ECMWF IFS
shows suggestive consistency in shape and rate-limiting stage (Fig.~\ref{fig:committor}),
providing qualitative evidence that the bottleneck structure is not an artefact of the
neural emulator's dynamics.

The same approach should also apply to other atmospheric rare events, including
extreme heatwaves, sudden stratospheric warmings, and blocking onset, and to
physics-based models when the ensemble cost is manageable.

\appendix
\section{State-$B$ Threshold Sensitivity}
\label{sec:si_threshold}

The choice of $\lambda_B = 975$~hPa was determined by the geographic and physical
character of state-$B$ arrivals, not solely by computational considerations.
Complete simulations were first carried out with $\lambda_B = 965$~hPa.
At that threshold, state-$B$ arrivals were disproportionately located at latitudes
$>40^\circ$N, with many reaching state~$B$ via extratropical deepening, including
cases analogous to Fiona's eventual high-latitude track, rather than through
tropical intensification.
Although the CPS warm-core filter rejects clearly extratropical trajectories,
at 965~hPa the filter struggled to cleanly separate borderline recurving storms
from genuine tropical development, leading to state-$B$ arrival distributions
inconsistent with observed Atlantic genesis climatology.

Raising $\lambda_B$ to 975~hPa substantially shifted state-$B$ arrivals into
the tropical and subtropical Atlantic, Caribbean, and Gulf of Mexico
(Fig.~\ref{fig:stateb_geo}), consistent with where September Atlantic storms
actually develop and track.
At this threshold the CPS filter operates more reliably, and the geographic
distribution of arrivals matches physical expectations for the 2022 season.
A secondary benefit was that $P_4$ values, which had collapsed below 0.05 across
most ICs at 965~hPa (concentrating nearly all variance in the final interface),
distributed more uniformly at 975~hPa, maintaining the well-conditioned range
targeted by the interface placement algorithm.

The shooting interface values (1000.0, 987.2, 984.7, 981.2~hPa) were optimised
using the 965~hPa simulations as a starting point: the crossing statistics from
those longer runs informed the optimal interface spacing for the 975~hPa state~$B$,
and the final values were rounded to physically interpretable pressure levels.

The sensitivity of reported genesis \textit{rates} to threshold choice is expected:
$k(\lambda_B = 975)$ and $k(\lambda_B = 965)$ differ by roughly one order of magnitude,
consistent with the additional conditional probability factor for the extra interface.
The key methodological properties (self-consistency, seasonal cycle structure,
and bottleneck diagnostics) are qualitatively robust across thresholds.
For physics-based models at finer resolution, a lower $\lambda_B$ (e.g., 960~hPa)
may be more appropriate, and the interface placement should be re-optimised accordingly.

\section{Convergence of FFS Rate Estimates}
\label{sec:convergence}

\begin{figure}[h]
\centering
\includegraphics[width=\textwidth]{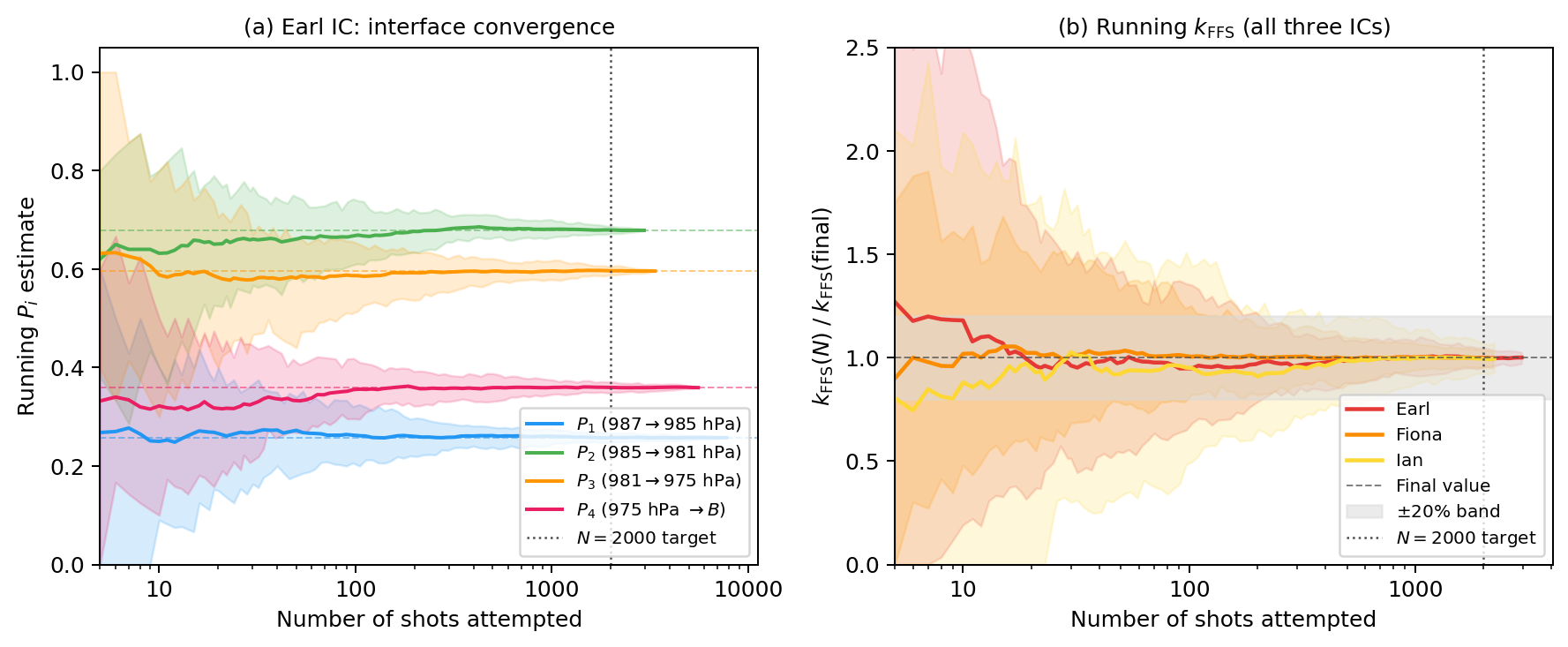}
\caption{Convergence of FFS estimates as a function of the number of shooting
attempts per interface.
(a) Running $P_i(N)$ for the four shooting interfaces of the Earl IC (mean and
10th--90th percentile across 50 random shot-orderings); dashed horizontal lines
mark the final $P_i$ values.
(b) Running $k_\mathrm{FFS}(N)$ normalised by the final value for all three
case-study ICs; gray shading spans $\pm 20\%$; dotted vertical line marks the
$N = 2000$ target.}
\label{fig:convergence}
\end{figure}

We target $N = 2000$ successful crossing attempts per interface to ensure stable
$P_i$ estimates and a well-converged genesis rate $k_\mathrm{FFS}$.
Figure~\ref{fig:convergence} demonstrates this convergence using the actual
shot-outcome logs from the three case-study initial conditions.

Panel~(a) shows the running $P_i$ estimate for each of the four shooting interfaces
of the Earl IC as a function of the number of shots attempted.
Each curve is the mean across 50 random permutations of the shot-ordering, with the
shaded envelope spanning the 10th--90th percentile range.
All four $P_i$ values stabilize well before $N = 2000$ (dotted vertical line), with
the envelope width falling below $\pm 0.02$ for the three lower interfaces and
$\pm 0.04$ for $P_4$ (the most uncertain, reflecting the near-threshold CPS filter).

Panel~(b) shows the running ratio $k_\mathrm{FFS}(N) / k_\mathrm{FFS}(\mathrm{final})$
for all three case-study ICs.
By $N \approx 500$ the estimates enter the $\pm 20\%$ band (gray shading), and by
$N = 2000$ they are within $\pm 5\%$ of their final values for all three storms.
The convergence behaviour is qualitatively similar across ICs despite the
approximately tenfold range in $k_\mathrm{FFS}$.

\section{IFS Ensemble Processing}
\label{sec:si_ifs}

The IFS quantities in Figures~\ref{fig:rates} and \ref{fig:committor} derive from the
ECMWF IFS 50-member ensemble archived by WeatherBench2 \citep{rasp2024weatherbench2}
at 1.5$^\circ$ output grid spacing, initialized at the same 98 initial-condition times
used for the FFS calculations.
For each ensemble member we apply the same vortex tracker described in Section~2.3 to
the MSLP field: Gaussian smoothing ($\sigma = 1.5$ grid points) locates local minima,
existing tracks are updated by nearest-neighbor matching within $12^\circ$, unmatched
low-latitude minima over land are suppressed as terrain artifacts, and a track is
retained for up to two consecutive timesteps of failed matching.
Systems already present at initialization are excluded, so only newly formed
disturbances contribute. New tracks are not seeded beyond the 15-day forecast window.
The $\lambda_0$ crossing rate is the number of crossings accumulated across members
divided by the total tracked time.
State-$B$ arrivals ($Q \leq 975$~hPa) provide the brute-force genesis rate with
Poisson uncertainties, and upper bounds are reported for initial conditions with
zero events.
The IFS committor at each interface is estimated as the fraction of tracked storms
crossing $\lambda_i$ that subsequently reach state~$B$ within the forecast window,
averaged across initial conditions.
Note that the IFS processing applies neither the decorrelated-crossing requirement
nor the CPS warm-core screen, so the comparison with the FFS quantities is
qualitative only (Section~\ref{sec:results_val}).

\section*{Acknowledgments}
This material is based upon work supported by the NSF National Center for Atmospheric
Research, which is a major facility sponsored by the U.S. National Science Foundation
under Cooperative Agreement No.\ 1852977.
This research has also been supported by NSF Grant No.\ RISE-2019758.
We would like to acknowledge high-performance computing support from Derecho and Casper
\citep{Cheyenne} provided by the Computational and Information Systems Laboratory,
NCAR, and sponsored by the National Science Foundation.

\section*{Data Availability Statement}
The FFS driver code, vortex tracker, and contamination screen are available at
\url{https://github.com/NCAR/miles-tails}.
Note that \texttt{config/ffs.yml} in the repository reflects a development-phase
parameter set. The operative interface values and thresholds are those stated in
Section~2.1 of this paper.
The FFS simulation output (rate statistics, conditional crossing probabilities, and
reactive pathway summaries for all 98 initial conditions) will be archived at the
NCAR Research Data Archive upon acceptance.
ERA5 reanalysis initial conditions \citep{hersbach2020era5} are available from the
Copernicus Climate Data Store (\url{https://cds.climate.copernicus.eu}).
IFS ensemble data are available through WeatherBench2 \citep{rasp2024weatherbench2}.
SDL-WXFormer model weights and details on the broader CREDIT framework are available at
\url{https://github.com/NCAR/miles-credit}.

\bibliographystyle{plainnat}
\bibliography{references}

@article{allen2005sampling,
  author  = {Allen, R.~J. and Warren, P.~B. and ten Wolde, P.~R.},
  title   = {Sampling rare switching events in biochemical networks},
  journal = {Physical Review Letters},
  volume  = {94},
  pages   = {018104},
  year    = {2005},
  doi     = {10.1103/PhysRevLett.94.018104}
}

@ARTICLE{allen2006forward,
  title     = "Simulating rare events in equilibrium or nonequilibrium
               stochastic systems",
  author    = "Allen, Rosalind J and Frenkel, Daan and ten Wolde, Pieter Rein",
  journal   = "J. Chem. Phys.",
  publisher = "AIP Publishing",
  volume    =  124,
  number    =  2,
  pages     =  024102,
  month     =  jan,
  year      =  2006,
  url       = "http://dx.doi.org/10.1063/1.2140273",
  doi       = "10.1063/1.2140273",
  language  = "en"
}

@article{allen2009forward,
  author  = {Allen, R.~J. and Valeriani, C. and ten Wolde, P.~R.},
  title   = {Forward flux sampling for rare event simulations},
  journal = {J. Phys.: Condens. Matter},
  volume  = {21},
  pages   = {463102},
  year    = {2009},
  doi     = {10.1088/0953-8984/21/46/463102}
}

@article{valeriani2007computing,
  author  = {Valeriani, C. and Allen, R.~J. and Morelli, M.~J. and Frenkel, D. and ten Wolde, P.~R.},
  title   = {Computing stationary distributions in equilibrium and nonequilibrium systems
             with forward flux sampling},
  journal = {J. Chem. Phys.},
  volume  = {127},
  pages   = {114109},
  year    = {2007},
  doi     = {10.1063/1.2767625}
}

@article{schreck2015dna,
  author  = {Schreck, J.~S. and Ouldridge, T.~E. and Romano, F. and \v{S}ulc, P. and Shaw, L.~P. and Louis, A.~A. and Doye, J.~P.~K.},
  title   = {{DNA} hairpins destabilize duplexes primarily by promoting melting rather than by inhibiting hybridization},
  journal = {Nucleic Acids Res.},
  volume  = {43},
  pages   = {6181--6190},
  year    = {2015},
  doi     = {10.1093/nar/gkv582}
}

@article{bolhuis2002transition,
  author  = {Bolhuis, P.~G. and Chandler, D. and Dellago, C. and Geissler, P.~L.},
  title   = {Transition path sampling: Throwing ropes over rough mountain passes,
             in the dark},
  journal = {Annu. Rev. Phys. Chem.},
  volume  = {53},
  pages   = {291--318},
  year    = {2002},
  doi     = {10.1146/annurev.physchem.53.082301.113146}
}

@ARTICLE{Khoo2019,
  title     = "Solving for high-dimensional committor functions using artificial
               neural networks",
  author    = "Khoo, Yuehaw and Lu, Jianfeng and Ying, Lexing",
  journal   = "Res. Math. Sci.",
  publisher = "Springer Science and Business Media LLC",
  volume    =  6,
  number    =  1,
  pages     =  1,
  month     =  mar,
  year      =  2019,
  doi       = "10.1007/s40687-018-0160-2",
  language  = "en"
}

@article{cerou2007adaptive,
  author  = {C{\'e}rou, F. and Guyader, A.},
  title   = {Adaptive multilevel splitting for rare event analysis},
  journal = {Stoch. Anal. Appl.},
  volume  = {25},
  pages   = {417--443},
  year    = {2007},
  doi     = {10.1080/07362990601139628}
}

@article{ragone2018computation,
  author  = {Ragone, F. and Wouters, J. and Bouchet, F.},
  title   = {Computation of extreme heat waves in climate models using a large
             deviation algorithm},
  journal = {Proc. Natl. Acad. Sci.},
  volume  = {115},
  pages   = {24--29},
  year    = {2018},
  doi     = {10.1073/pnas.1712645115}
}

@article{webber2019practical,
  author  = {Webber, R.~J. and Plotkin, D.~A. and O'Neill, M.~E. and
             Abbot, D.~S. and Weare, J.},
  title   = {Practical rare event sampling for extreme mesoscale weather},
  journal = {Chaos},
  volume  = {29},
  pages   = {053109},
  year    = {2019},
  doi     = {10.1063/1.5081461}
}

@ARTICLE{finkel2024bringing,
  title     = "Bringing statistics to storylines: Rare event sampling for
               sudden, transient extreme events",
  author    = "Finkel, Justin and O'Gorman, Paul A",
  journal   = "J. Adv. Model. Earth Syst.",
  publisher = "American Geophysical Union (AGU)",
  volume    =  16,
  number    =  6,
  pages     = "e2024MS004264",
  month     =  jun,
  year      =  2024,
  url       = "http://dx.doi.org/10.1029/2024MS004264",
  doi       = "10.1029/2024ms004264",
  language  = "en"
}

@article{plotkin2019maximizing,
  author  = {Plotkin, D.~A. and Webber, R.~J. and O'Neill, M.~E. and
             Weare, J. and Abbot, D.~S.},
  title   = {Maximizing simulated tropical cyclone intensity with action minimization},
  journal = {J. Adv. Model. Earth Syst.},
  volume  = {11},
  pages   = {863--891},
  year    = {2019},
  doi     = {10.1029/2018MS001419}
}

@article{galfi2019large,
  author  = {G{\'a}lfi, V.~M. and Lucarini, V. and Wouters, J.},
  title   = {A large deviation theory-based analysis of heat waves and cold spells
             in a simplified model of the general circulation of the atmosphere},
  journal = {J. Stat. Mech.: Theory Exp.},
  volume  = {2019},
  pages   = {033404},
  year    = {2019},
  doi     = {10.1088/1742-5468/ab02e8}
}

@article{touchette2009large,
  author  = {Touchette, H.},
  title   = {The large deviation approach to statistical mechanics},
  journal = {Phys. Rep.},
  volume  = {478},
  pages   = {1--69},
  year    = {2009},
  doi     = {10.1016/j.physrep.2009.05.002}
}

@ARTICLE{lam2023graphcast,
  title    = "Learning skillful medium-range global weather forecasting",
  author   = "Lam, Remi and Sanchez-Gonzalez, Alvaro and Willson, Matthew and
              Wirnsberger, Peter and Fortunato, Meire and Alet, Ferran and
              Ravuri, Suman and Ewalds, Timo and Eaton-Rosen, Zach and Hu,
              Weihua and Merose, Alexander and Hoyer, Stephan and Holland,
              George and Vinyals, Oriol and Stott, Jacklynn and Pritzel,
              Alexander and Mohamed, Shakir and Battaglia, Peter",
  journal  = "Science",
  volume   =  382,
  number   =  6677,
  pages    = "1416--1421",
  month    =  dec,
  year     =  2023,
  doi      = "10.1126/science.adi2336",
  language = "en"
}

@ARTICLE{bi2023pangu,
  title    = "Accurate medium-range global weather forecasting with {3D} neural
              networks",
  author   = "Bi, Kaifeng and Xie, Lingxi and Zhang, Hengheng and Chen, Xin and
              Gu, Xiaotao and Tian, Qi",
  journal  = "Nature",
  volume   =  619,
  number   =  7970,
  pages    = "533--538",
  month    =  jul,
  year     =  2023,
  url      = "http://dx.doi.org/10.1038/s41586-023-06185-3",
  doi      = "10.1038/s41586-023-06185-3",
  language = "en",
}

@ARTICLE{pathak2022fourcastnet,
  title         = "{FourCastNet}: A Global Data-driven High-resolution Weather
                   Model using Adaptive Fourier Neural Operators",
  author        = "Pathak, Jaideep and Subramanian, Shashank and Harrington,
                   Peter and Raja, Sanjeev and Chattopadhyay, Ashesh and
                   Mardani, Morteza and Kurth, Thorsten and Hall, David and Li,
                   Zongyi and Azizzadenesheli, Kamyar and Hassanzadeh, Pedram
                   and Kashinath, Karthik and Anandkumar, Animashree",
  journal       = "arXiv [physics.ao-ph]",
  month         =  feb,
  year          =  2022,
  doi           =  {10.48550/arXiv.2202.11214},
  archivePrefix = "arXiv",
  primaryClass  = "physics.ao-ph",
  eprint        = "2202.11214"
}

@article{schreck2025sdl,
  author  = {Schreck, J.~S. and Chapman, W.~E. and Becker, C. and Gagne, D.~J. and
             Kimpara, D. and Cherukuru, N. and Berner, J. and Mayer, K.~J. and Sobhani, N.},
  title   = {Controllable probabilistic forecasting with stochastic decomposition layers},
  journal = {arXiv:2512.18815},
  year    = {2025},
  doi     = {10.48550/arXiv.2512.18815}
}

@inbook{ifs_ens_docs,
  author = {ECMWF},
  title = {IFS Documentation CY48R1 - Part V: Ensemble Prediction System},
  year = {2023},
  journal = {IFS Documentation CY48R1},
  number = {5},
  month = {06/2023},
  publisher = {ECMWF},
  doi = {10.21957/e529074162},
}

@misc{Cheyenne,
  author    = {{Computational and Information Systems Laboratory}},
  title     = {Derecho: {HPE System} ({NCAR Community Computing})},
  publisher = {NSF National Center for Atmospheric Research},
  year      = {2023},
  url       = "https://doi.org/10.5065/D6RX99HX"
}

@article{hersbach2020era5,
  author  = {Hersbach, H. and Bell, B. and Berrisford, P. and Hirahara, S. and
             Hor{\'a}nyi, A. and Mu{\~n}oz-Sabater, J. and Nicolas, J. and
             Peubey, C. and Radu, R. and Schepers, D. and Simmons, A. and
             Soci, C. and Abdalla, S. and Abellan, X. and Balsamo, G. and
             Bechtold, P. and Biavati, G. and Bidlot, J. and Bonavita, M. and
             De Chiara, G. and Dahlgren, P. and Dee, D. and Diamantakis, M. and
             Dragani, R. and Flemming, J. and Forbes, R. and Fuentes, M. and
             Geer, A. and Haimberger, L. and Healy, S. and Hogan, R.~J. and
             H{\'o}lm, E. and Janiskov{\'a}, M. and Keeley, S. and Laloyaux, P. and
             Lopez, P. and Lupu, C. and Radnoti, G. and de Rosnay, P. and
             Rozum, I. and Vamborg, F. and Villaume, S. and Th{\'e}paut, J.-N.},
  title   = {The {ERA5} global reanalysis},
  journal = {Quart. J. Roy. Meteor. Soc.},
  volume  = {146},
  pages   = {1999--2049},
  year    = {2020},
  doi     = {10.1002/qj.3803}
}

@article{Frank1970,
  author  = {Frank, N.~L.},
  title   = {Atlantic tropical systems of 1969},
  journal = {Mon. Wea. Rev.},
  volume  = {98},
  pages   = {307--314},
  year    = {1970},
  doi     = {10.1175/1520-0493(1970)098<0307:ATSO>2.3.CO;2}
}

@article{gray1968global,
  author  = {Gray, W.~M.},
  title   = {Global view of the origin of tropical disturbances and storms},
  journal = {Mon. Wea. Rev.},
  volume  = {96},
  pages   = {669--700},
  year    = {1968},
  doi     = {10.1175/1520-0493(1968)096%3C0669:GVOTOO%3E2.0.CO;2}
}

@article{Landsea1998,
  author  = {Landsea, C.~W.},
  title   = {Atlantic basin hurricanes: Indices of climatic changes},
  journal = {Climatic Change},
  volume  = {42},
  pages   = {89--129},
  year    = {1998},
  doi     = {10.1023/A:1005416332322}
}

@article{Halperin2013,
  author  = {Halperin, D.~J. and Fuelberg, H.~E. and Hart, R.~E. and
             Cossuth, J.~H. and Sura, P. and Pasch, R.~J.},
  title   = {An evaluation of tropical cyclone genesis forecasts from global
             numerical models},
  journal = {Wea. Forecasting},
  volume  = {28},
  pages   = {1423--1445},
  year    = {2013},
  doi     = {10.1175/WAF-D-13-00008.1}
}

@article{hart2003cyclone,
  author  = {Hart, R.~E.},
  title   = {A cyclone phase space derived from thermal wind and thermal asymmetry},
  journal = {Mon. Wea. Rev.},
  volume  = {131},
  pages   = {585--616},
  year    = {2003},
  doi     = {10.1175/1520-0493(2003)131%3C0585:ACPSDF%3E2.0.CO;2}
}

@article{Titley2000,
  author  = {Titley, D.~W. and Elsberry, R.~L.},
  title   = {Large intensity changes in tropical cyclones: A climatology and case
             studies for the western {North Pacific}},
  journal = {Mon. Wea. Rev.},
  volume  = {128},
  pages   = {92--111},
  year    = {2000},
  doi     = {10.1175/1520-0493(2000)128%3C3556:LICITC%3E2.0.CO;2}
}

@article{knaff2007improved,
  author  = {Knaff, J.~A. and Zehr, R.~M.},
  title   = {Reexamination of tropical cyclone pressure--wind relationships},
  journal = {Wea. Forecasting},
  volume  = {22},
  pages   = {71--88},
  year    = {2007},
  doi     = {10.1175/WAF965.1}
}

@article{lorenz1969atmospheric,
  author  = {Lorenz, E.~N.},
  title   = {The predictability of a flow which possesses many scales of motion},
  journal = {Tellus},
  volume  = {21},
  pages   = {289--307},
  year    = {1969},
  doi     = {10.1111/j.2153-3490.1969.tb00444.x}
}

@Article{heikenfeld2019tobac,
AUTHOR = {Heikenfeld, M. and Marinescu, P. J. and Christensen, M. and Watson-Parris, D. and Senf, F. and van den Heever, S. C. and Stier, P.},
TITLE = {tobac 1.2: towards a flexible framework for tracking and analysis of clouds in diverse datasets},
JOURNAL = {Geoscientific Model Development},
VOLUME = {12},
YEAR = {2019},
NUMBER = {11},
PAGES = {4551--4570},
DOI = {10.5194/gmd-12-4551-2019}
}

@article{bister1998dissipative,
  author  = {Bister, M. and Emanuel, K. A.},
  title   = {Dissipative heating and hurricane intensity},
  journal = {Meteor.\ Atmos.\ Phys.},
  volume  = {65},
  pages   = {233--240},
  year    = {1998},
  doi     = {10.1007/BF01030791}
}

@article{camargo2007use,
  author  = {Camargo, S. J. and Emanuel, K. A. and Sobel, A. H.},
  title   = {Use of a genesis potential index to diagnose {ENSO} effects on tropical cyclone genesis},
  journal = {J.\ Climate},
  volume  = {20},
  pages   = {4819--4834},
  year    = {2007},
  doi     = {10.1175/JCLI4282.1}
}

@ARTICLE{rasp2024weatherbench2,
  title     = "{WeatherBench} 2: A benchmark for the next generation of
               data‐driven global weather models",
  author    = "Rasp, Stephan and Hoyer, Stephan and Merose, Alexander and
               Langmore, Ian and Battaglia, Peter and Russell, Tyler and
               Sanchez-Gonzalez, Alvaro and Yang, Vivian and Carver, Rob and
               Agrawal, Shreya and Chantry, Matthew and Ben Bouallegue, Zied and
               Dueben, Peter and Bromberg, Carla and Sisk, Jared and Barrington,
               Luke and Bell, Aaron and Sha, Fei",
  journal   = "J. Adv. Model. Earth Syst.",
  publisher = "American Geophysical Union (AGU)",
  volume    =  16,
  number    =  6,
  pages     = "e2023MS004019",
  month     =  jun,
  year      =  2024,
  doi       = "10.1029/2023ms004019",
  language  = "en"
}

@article{hussain2020studying,
  author  = {Hussain, S. and Haji-Akbari, A.},
  title   = {Studying rare events using forward-flux sampling: Recent breakthroughs and future outlook},
  journal = {J.\ Chem.\ Phys.},
  volume  = {152},
  pages   = {060901},
  year    = {2020},
  doi     = {10.1063/1.5127780}
}

@article{lancelin2025ai,
  author  = {Lancelin, A. and Wikner, A. and Dubus, L. and {Le Priol}, C. and Abbot, D.~S. and Bouchet, F. and Hassanzadeh, P. and Weare, J.},
  title   = {{AI}-boosted rare event sampling to characterize extreme weather},
  journal = {arXiv preprint arXiv:2510.27066},
  year    = {2025},
  eprint  = {2510.27066},
  archivePrefix = {arXiv},
  doi = {10.48550/arXiv.2510.27066}
}

@article{finkel2023revealing,
  author  = {Finkel, J. and Gerber, E.~P. and Abbot, D.~S. and Weare, J.},
  title   = {Revealing the statistics of extreme events hidden in short weather forecast data},
  journal = {AGU Adv.},
  volume  = {4},
  pages   = {e2023AV000881},
  year    = {2023},
  doi     = {10.1029/2023AV000881}
}

@article{zhang2026physics,
  author  = {Zhang, Z. and Fischer, E. and Zscheischler, J. and Engelke, S.},
  title   = {Physics-based models outperform {AI} weather forecasts of record-breaking extremes},
  journal = {Sci.\ Adv.},
  year    = {2026},
  doi     = {10.1126/sciadv.aec1433}
}

@article{manshausen2026extreme,
  author  = {Manshausen, P. and Brenowitz, N. and Berner, J. and Kashinath, K. and Pritchard, M.},
  title   = {Towards accurate extreme event likelihoods from diffusion model climate emulators},
  journal = {arXiv preprint arXiv:2605.03802},
  year    = {2026},
  eprint  = {2605.03802},
  archivePrefix = {arXiv},
  doi = {10.48550/arXiv.2605.03802}
}

\end{document}